\documentclass[journal]{IEEEtran}
\ifCLASSINFOpdf
\else
 \usepackage[dvips]{graphicx}
 \usepackage{color}
\fi
\usepackage[cmex10]{amsmath}
\usepackage{amssymb}
\usepackage[tight,footnotesize]{subfigure}
\begin{document}
\title{Massive MIMO Relaying with Linear Precoding in Correlated Channels under Limited Feedback}

\author{$\mbox{Yang~Liu}${$^{^\dagger}$}, Zhiguo~Ding, Jia~Shi, Weiwei~Yang and Ping~Zhong% <-this % stops a space
\thanks{$\mbox{Y.~Liu}${$^{^\dagger}$}(e-mail: {\tt ly@bupt.edu.cn}) is with the School of Electronic Engineering, Beijing University of Posts and Telecommunications, Beijing 100876, China.}
\thanks{Z.~Ding is with the School of Computing and Communications, Lancaster University, Lancaster, LA1 4WA, UK.}
\thanks{J.~Shi is with the University of Surrey, Guildford, GU2 7XH, UK.}
\thanks{W.~Yang is with the College of Communications Engineering, PLA University of Science and Technology, Nanjing, China.}
\thanks{P.~Zhong is with the School of Mathematics and Statistics, University of Wuhan, Wuhan 430074, China. P.~Zhong is also with the Department of Pure Mathematics, University of Waterloo, Waterloo, Ontario, Canada N2L 3G1.}}

\maketitle

\begin{abstract}
In this paper we study on a massive MIMO relay system with linear precoding under the conditions of imperfect channel state information at the transmitter (CSIT) and per-user channel transmit correlation. In our system the source-relay channels are massive multiple-input multiple-output (MIMO) ones and the relay-destination channels are massive multiple-input single-output (MISO) ones. Large random matrix theory (RMT) is used to derive a deterministic equivalent of the signal-to-interference-plus-noise ratio (SINR) at each user in massive MIMO amplify-forward and decode-forward (M-MIMO-ADF) relaying with regularized zero-forcing (RZF) precoding, as the number of transmit antennas and users $M,K \to \infty $ and $M \gg K$. In this paper we obtain a closed-form expression for the deterministic equivalent of ${\mathbf{h}}_k^H{{\mathbf{\hat W}}_l}{{\mathbf{\hat h}}_k}$, and we give two theorems and a corollary to derive the deterministic equivalent of the SINR at each user. Simulation results show that the deterministic equivalent of the SINR at each user in M-MIMO-ADF relaying and the results of Theorem 1, Theorem 2, Proposition 1 and Corollary 1 are accurate.
\end{abstract}


% \fi
\IEEEpeerreviewmaketitle

\section{Introduction}
\IEEEPARstart{I}{n} recent years massive multiple-input multiple-output (MIMO) antenna system has become a key pillar in the research of the 5th generation (5G) wireless communication technologies. As compared to conventional MIMO systems, massive MIMO systems can bring several orders of magnitude gain to spectral and energy efficiency [1]-[4]. It has been proved that the sum capacity of multi-user massive MIMO (MU-M-MIMO) channels can be greatly enhanced as compared to that of single user massive MIMO (SU-M-MIMO) channels. However, MU-M-MIMO suffers from pilot contamination (i.e., residual interference which is caused by the reuse of pilot sequences in adjacent cells) [5], inter-user interference at receivers [6][7], and imperfect channel state information (CSI) acquisition [8][9], etc.

To mitigate inter-user interferences, appropriate precoding is needed at transmitters. As we know, dirty-paper coding (DPC) is a capacity achieving precoding strategy for the Gaussian MIMO broadcast channels (MIMO-BC) [10][11]. However, DPC precoder is nonlinear and it is much more complex to be implemented for practical use. For a massive MIMO system with large number of antennas, low-complexity linear precoding such as zero-forcing (ZF) [11][12] and regularized zero-forcing (RZF) can achieve near-optimal rate performance and thus is more suitable for practical use. On the other hand, precoding for massive MIMO systems requires accurate instantaneous channel state information at the transmitter (CSIT) which is untenable in practice, and the inaccuracy of CSIT will seriously affect the performance of massive MIMO systems. For this reason, many research works study on massive MIMO systems with imperfect CSIT [13][14][21][22][26].

Besides direct transmission, many research works also study on the application of massive MIMO to cooperative relay systems, e.g., there have been some works on multi-way massive MIMO relay systems [15][16]. These massive MIMO relay systems can offer the benefits of both massive MIMO and multi-way relay systems, thus are expected to achieve very high spectral efficiency. One of the disadvantages of massive MIMO systems is the increased hardware and software complexity, which can be reduced by antenna selection at base station (BS) or relays. Antenna selection has been a key topic in the field of massive MIMO research in the past years, the use of antenna selection in transmission or reception also has the benefits of improving power efficiency and improving system performance [17]-[19].

For a massive MIMO system with number of antennas $M \to \infty $, if partial CSI such as statistical CSI and channel correlation matrices are available, then a large system analysis can be carried out and the signal-to-interference-plus-noise ratio (SINR) can be approximated by a deterministic equivalent. Hochwald \emph{et al}. [20] were the first to study on large system analysis with $M,K \to \infty $ ($M$ and $K$ denote the number of antennas and users, respectively) and finite ratio for linear precoding. Wagner \emph{et al}. [21] studied large system analysis on the sum rate performance of ZF and RZF precoding for large multiple-input single-output (MISO) broadcast systems under the conditions of imperfect CSIT and per-user channel transmit correlation. Besides, [22][23] analyzed a rate-splitting (RS) and a hierarchical-rate-splitting (HRS) scheme for massive MIMO with imperfect CSIT in a large-scale array regime. [24] studied a non-regenerative massive MIMO non-orthogonal multiple access (NOMA) relay system, in which SU-M-MIMO is applied between the BS and the relay while MU-M-MIMO is applied between the relay and the users. Large system analysis for massive MIMO can also be found in [7][13][25][26][27].

In this paper, we study on a massive MIMO relay system with linear precoding under limited feedback, i.e., a BS equipped with $M$ ($M \to \infty $) antennas communicates $K$ single-antenna non-cooperative receivers with the help of a relay equipped with $M+K$ antennas. We consider RZF precoding under the conditions of imperfect CSIT and per-user channel transmit correlation, and the relay schemes are based on half-duplex amplify-forward (AF) or decode-forward (DF) mode. The main contributions of this paper can be summarized as follows:
\begin{itemize}
  \item For massive MIMO AF and DF (M-MIMO-ADF) relaying with RZF precoding under the conditions of imperfect CSIT, per-user channel correlation, and that the source-relay channels are massive MIMO ones rather than massive MISO ones, we approximates the SINR at each user by a deterministic equivalent. This paper is one of the few works involving deterministic approximations for precoding in massive MIMO channels (rather than massive MISO channels). Compared with SU-M-MIMO or MU-M-MIMO method, our method can make a real system easy to be synchronized.
  \item Under the conditions of imperfect CSIT and per-user channel transmit correlation, we give two theorems and a corollary to derive the deterministic equivalent of the SINR at each user; and different from the iterative method in [21] to obtain the deterministic equivalent of ${\mathbf{h}}_k^H{{\mathbf{\hat W}}_l}{{\mathbf{\hat h}}_k}$, we obtain a closed-form expression for the deterministic equivalent of ${\mathbf{h}}_k^H{{\mathbf{\hat W}}_l}{{\mathbf{\hat h}}_k}$.
  \item Simulation results match well with theoretical analysis. It shows that the deterministic equivalent of the SINR at each user in M-MIMO-ADF relaying and the results of Theorem 1, Theorem 2, Proposition 1 and Corollary 1 are accurate.
\end{itemize}

The rest of this paper is organized as follows. System model and scheme description is introduced in Section II. Section III derives the deterministic equivalent of the SINR at each user under the conditions of imperfect CSIT and common correlation. Simulation results are given in Section IV. Finally, conclusions are summarized in Section V.

The proofs for Theorem 1, Theorem 2, Proposition 1 and Corollary 1 are presented in Appendix A, B, and C respectively, and some lemmas which are used in our proofs are collected in Appendix D.

$Notation:$ In the following, boldface lower-case and upper-case characters denote vectors and matrices, respectively. The operators ${\left(  \cdot  \right)^H}$, ${\left(  \cdot  \right)^T}$, tr$\left(  \cdot  \right)$ and $E\left[  \cdot  \right]$ denote conjugate transpose, transpose, trace and expectation, respectively. ${{\mathbf{I}}_M}$ denotes the $M \times M$ identity matrix. $\log \left(  \cdot  \right)$ is the natural logarithm.

%I wish you the best of success.



\section{System Model and Scheme Description}
\subsection{System Model}
Consider a downlink transmission scenario with one BS transmitting data to multiple users with the help of a single relay. The BS sends independent user symbols ${\mathbf{s}} = {\left[ {{s_1}, \cdots ,{s_K}} \right]^T}$ to $K$ users, where the data ${s_k}$ is for the $k$-th user ($k \in \left\{ {1,2, \cdots ,K} \right\}$). The BS is equipped with $M$ antennas, each user is equipped with a single antenna. It is assumed there is no direct link between the BS and the users. The relay works in half-duplex mode and is equipped with $M+K$ antennas, where $M$ is the number of transmit antennas, and $K$ is the number of receive antennas of the relay. In this paper a large system analysis is carried out with $M \to \infty $, $K \to \infty $ for $M \gg K$. Here $M \to \infty $ means that the BS and the relay are equipped with hundreds of transmit antennas. Besides, $M \ge K$ is assumed to make user scheduling not have to be analyzed. Please note that a massive MIMO relay system may be sensitive to delay, e.g., massive MIMO relaying in a vehicle network. As compared to decoding with all the receive antennas of the relay, the first hop decoding with $K$ receive antennas of the relay is a more simple task and may be accomplished with less delay. Besides, such a system setup (that the relay uses small number ($K$) of antennas to receive signals, and then uses many ($M$) antennas to transmit signals) is a balanced setup for the first (BS$\rightarrow$relay) and the second hop (relay$\rightarrow$users) of the relay system, which can achieve both large sum rate and low delay.
\begin{figure}[!t]
\centering
\includegraphics[width=3.5in]{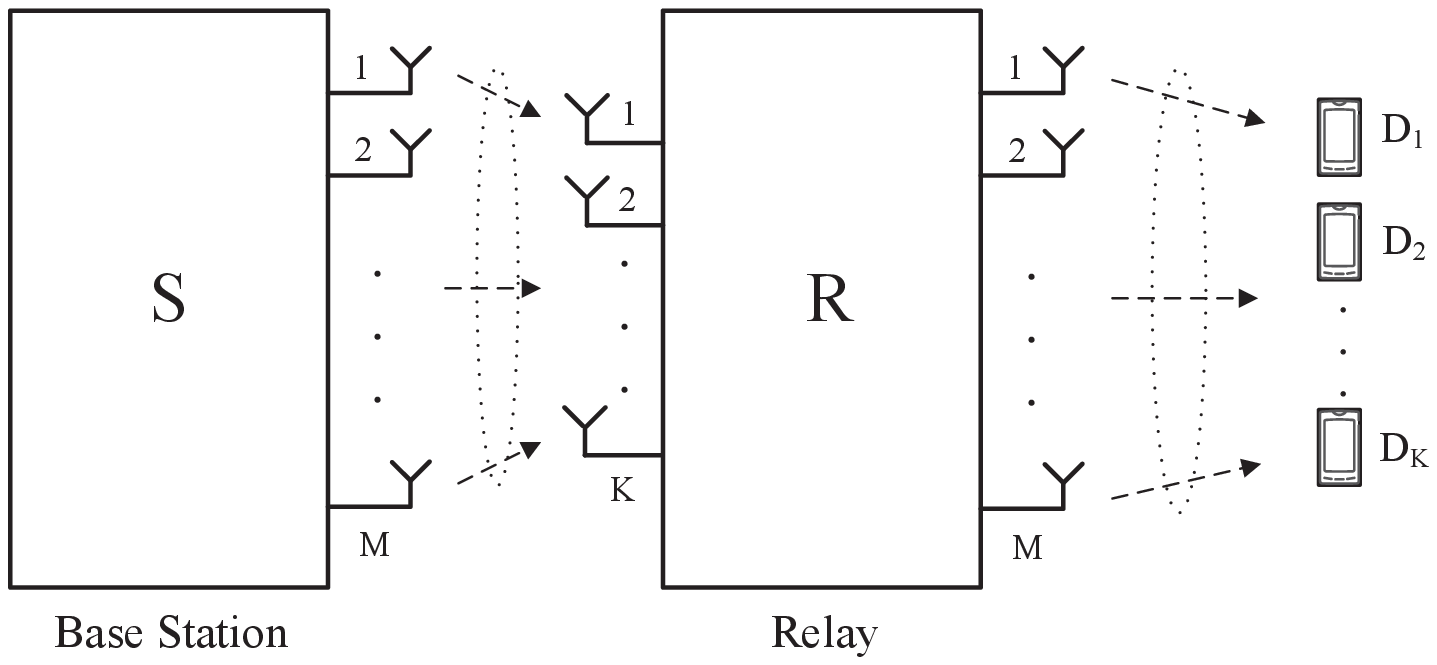}
% where an .eps filename suffix will be assumed under latex,
% and a .pdf suffix will be assumed for pdflatex; or what has been declared
% via \DeclareGraphicsExtensions.
\caption{System model.}
\label{Fig1}
\end{figure}

With a RZF precoding, the transmitted signal at the BS is
\begin{equation} \label{eq:1}
{{\mathbf{x}}_s} = \sum\limits_{k = 1}^K {\sqrt {{p_{s,k}}} {{\mathbf{g}}_{sr,k}}{s_k}}
\end{equation}
where ${{\mathbf{g}}_{sr,k}} \in {\mathbb{C}^M}$ denotes the precoding vector for a source-relay channel from the BS to the $k$-th receive antenna of the relay, ${p_{s,k}}$ denotes the transmit power for symbol $s_k$, $P$ is the transmit power constraint for the BS or the relay.

In this system each channel is modeled as
\begin{equation} \label{eq:2}
{{\mathbf{h}}_{ij,k}} = \sqrt M {\mathbf{\Theta }}_{ij,k}^{1/2}{{\mathbf{z}}_{ij,k}}
\end{equation}
where $ij \in \left\{ {sr,rd} \right\}$, $sr$ and $rd$ denotes a source-relay and a relay-destination channel respectively, $k \in \left\{ {1, \cdots ,K} \right\}$, ${{\mathbf{\Theta }}_{ij,k}}$ denotes the channel correlation matrix which is assumed to be slowly varying as compared to the channel coherence time and thus are supposed to be perfectly known to the transmitter, ${{\mathbf{z}}_{ij,k}}\sim \mathcal{C}\mathcal{N}\left( {0,{1 \mathord{\left/
 {\vphantom {1 M}} \right.
 \kern-\nulldelimiterspace} M}} \right)$ denotes the quasi-static independent and identically distributed (i.i.d.) fast fading channel vector. User $k$ is assumed to have only knowledge about ${{\mathbf{\Theta }}_{ij,k}}$. Moreover, only an imperfect estimated channel of ${{\mathbf{\hat h}}_{ij,k}}$ is assumed to be available at the transmitter and the relay which is modeled as
\begin{eqnarray} \label{eq:3}
  {{{\mathbf{\hat h}}}_{ij,k}} &\!\!\!=\!\!\!& \sqrt M {\mathbf{\Theta }}_{ij,k}^{1/2}\left( {\sqrt {1 - \tau _{ij,k}^2} {{\mathbf{z}}_{ij,k}} + {\tau _{ij,k}}{{\mathbf{q}}_{ij,k}}} \right) \hfill \nonumber \\
  &\!\!\!=\!\!\!& \sqrt M {\mathbf{\Theta }}_{ij,k}^{1/2}{{{\mathbf{\hat z}}}_{ij,k}} \hfill
\end{eqnarray}
where ${{\mathbf{\hat z}}_{ij,k}} \!=\! \sqrt {1 \!-\! \tau _{ij,k}^2} {{\mathbf{z}}_{ij,k}} + {\tau _{ij,k}}{{\mathbf{q}}_{ij,k}}$, ${{\mathbf{q}}_{ij,k}} \! \sim \! \mathcal{C}\mathcal{N}\left( {0,{1 \mathord{\left/
 {\vphantom {1 M}} \right.
 \kern-\nulldelimiterspace} M}} \right)$ is independent of ${{\mathbf{z}}_{ij,k}}$, ${\tau _{ij,k}}\in [0,1]$ reflects the accuracy of channel estimation.

\subsection{Scheme Description}
Next we introduce the transmission scheme. Let the RZF precoding vector be
\begin{equation} \label{eq:4}
{{\mathbf{g}}_{sr,k}} = {\xi _1}{{\mathbf{\hat W}}_1}{{\mathbf{\hat h}}_{sr,k}},\quad {{\mathbf{g}}_{rd,k}} = {\xi _2}{{\mathbf{\hat W}}_2}{{\mathbf{\hat h}}_{rd,k}},
\end{equation}
where ${{\mathbf{\hat W}}_1} = {\left( {{\mathbf{\hat H}}_1^H{{{\mathbf{\hat H}}}_1} + M{\alpha _1}{{\mathbf{I}}_M}} \right)^{ - 1}} \in {\mathbb{C}^{M \times M}}$, ${{\mathbf{\hat H}}_1} = {\left[ {{{{\mathbf{\hat h}}}_{sr,1}}, \cdots ,{{{\mathbf{\hat h}}}_{sr,K}}} \right]^H} \in {\mathbb{C}^{K \times M}}$, ${{\mathbf{\hat h}}_{sr,k}},{{\mathbf{\hat h}}_{rd,k}} \in {\mathbb{C}^M}$, ${{\mathbf{\hat W}}_2} = {\left( {{\mathbf{\hat H}}_2^H{{{\mathbf{\hat H}}}_2} + M{\alpha _2}{{\mathbf{I}}_M}} \right)^{ - 1}} \in {\mathbb{C}^{M \times M}}$, ${{\mathbf{\hat H}}_2} = {\left[ {{{{\mathbf{\hat h}}}_{rd,1}}, \cdots ,{{{\mathbf{\hat h}}}_{rd,K}}} \right]^H} \in {\mathbb{C}^{K \times M}}$, ${\xi _2} = {\xi _{AF}}$ for AF relaying, and ${\xi _2} = {\xi _{DF}}$ for DF relaying.

In the first time slot, the received signal at the relay is given by
\begin{eqnarray} \label{eq:5}
&&\!\!\!\!\!\!\!\!\!\!  {{\mathbf{y}}_{sr}} = {{\mathbf{H}}_{sr}}{{\mathbf{x}}_s} + {{\mathbf{n}}_r} \hfill \nonumber \\
   && = \left[ \begin{gathered}
  {\mathbf{h}}_{sr,1}^H  \\
   \vdots   \\
  {\mathbf{h}}_{sr,K}^H  \\
\end{gathered}  \right]\left[ {\sum\limits_{k = 1}^K {\sqrt {\frac{{G{p_{s,k}}}}{{r_{sr}^\alpha }}} {{\mathbf{g}}_{sr,k}}{s_k}} } \right] + \left[ \begin{gathered}
  {n_{r,1}}  \\
   \vdots   \\
  {n_{r,K}}  \\
\end{gathered}  \right] \hfill \nonumber \\
   && = \!\!\left[\!\! {\begin{array}{*{20}{c}}
  {{\mathbf{h}}_{sr,1}^H\sqrt {\frac{{G{p_{s,1}}}}{{r_{sr}^\alpha }}} {{\mathbf{g}}_{sr,1}}} \\
   \vdots  \\
  {{\mathbf{h}}_{sr,K}^H\sqrt {\frac{{G{p_{s,1}}}}{{r_{sr}^\alpha }}} {{\mathbf{g}}_{sr,1}}}
\end{array}} \!\!\right]\!{s_1} \!+\! \left[\!\! {\begin{array}{*{20}{c}}
  {{\mathbf{h}}_{sr,1}^H\sqrt {\frac{{G{p_{s,2}}}}{{r_{sr}^\alpha }}} {{\mathbf{g}}_{sr,2}}} \\
   \vdots  \\
  {{\mathbf{h}}_{sr,K}^H\sqrt {\frac{{G{p_{s,2}}}}{{r_{sr}^\alpha }}} {{\mathbf{g}}_{sr,2}}}
\end{array}} \!\!\right]\!{s_2} \nonumber \\
&& \quad +  \cdots  + \left[\!\! {\begin{array}{*{20}{c}}
  {{\mathbf{h}}_{sr,1}^H\sqrt {\frac{{G{p_{s,K}}}}{{r_{sr}^\alpha }}} {{\mathbf{g}}_{sr,K}}} \\
   \vdots  \\
  {{\mathbf{h}}_{sr,K}^H\sqrt {\frac{{G{p_{s,K}}}}{{r_{sr}^\alpha }}} {{\mathbf{g}}_{sr,K}}}
\end{array}} \!\!\right]\!{s_K} + \!\left[ \!\begin{gathered}
  {n_{r,1}}  \\
   \vdots   \\
  {n_{r,K}}  \\
\end{gathered}  \! \right] \hfill
\end{eqnarray}
where ${{\mathbf{H}}_{sr}} = {\left[ {{{\mathbf{h}}_{sr,1}}, \cdots ,{{\mathbf{h}}_{sr,K}}} \right]^H} \in {\mathbb{C}^{K \times M}}$ denotes the i.i.d. channel matrix from the BS to the relay, ${{\mathbf{h}}_{sr,k}} \in {\mathbb{C}^M}$ ($k \in \left\{ {1, \cdots ,K} \right\}$) denotes the i.i.d. channel vector from the BS to the $k$-th receive antenna of the relay; ${r_{sr}}$ is the distance from the BS to the relay; $\alpha$ is the path loss exponent, $G$ is a constant that incorporates the effects of path loss, antenna gain, antenna height, and other factors; ${{\mathbf{n}}_r}$ denotes the additive white Gaussian noise (AWGN) vector at the relay, ${n_{r,k}} \in \mathcal{C}\mathcal{N}\left( {0,{N_0}} \right)$ denotes the AWGN at the $k$-th receive antenna of the relay. Then the received signal at the $k$-th receive antenna of the relay is given by
\begin{equation} \label{eq:6}
{y_{sr,k}} = {\mathbf{h}}_{sr,k}^H\sum\limits_{n = 1}^K {\sqrt {\frac{{G{p_{s,n}}}}{{r_{sr}^\alpha }}} {{\mathbf{g}}_{sr,n}}{s_n}}  + {n_{r,k}}
\end{equation}

Before we move on to the next discussion, let's introduce the following theorem which is useful for the subsequent large system analysis.

\textit{Theorem 1:} Let ${\mathbf{\Phi }}\! = \!\left\{ {{{\mathbf{h}}_{sr,1}}, \cdots ,{{\mathbf{h}}_{sr,K}},{{\mathbf{h}}_{rd,1}}, \cdots ,{{\mathbf{h}}_{rd,K}}} \right\}$, ${\mathbf{\hat \Phi }}\! =\! \left\{ {{{{\mathbf{\hat h}}}_{sr,1}}, \cdots ,{{{\mathbf{\hat h}}}_{sr,K}},{{{\mathbf{\hat h}}}_{rd,1}}, \cdots ,{{{\mathbf{\hat h}}}_{rd,K}}} \right\}$, ${{\mathbf{h}}_k} \in {\mathbf{\Phi }}$, ${{\mathbf{\hat h}}_k} \in {\mathbf{\hat \Phi }}$, ${{\mathbf{\hat h}}_{k'}} \in {\mathbf{\hat \Phi }} - {{\mathbf{\hat h}}_k}$, then we have $\left| {{{\mathbf{h}}_k^H}{{{\mathbf{\hat W}}}_l}{{{\mathbf{\hat h}}}_k}} \right| \gg \left| {{{\mathbf{h}}_k^H}{{{\mathbf{\hat W}}}_l}{{{\mathbf{\hat h}}}_{k'}}} \right|$ for any $k \ne k'$, $M \to \infty $ and $M \gg K$, where $k,k' \in \left\{ {1, \cdots ,K} \right\}$ and $l \in \left\{ {1,2} \right\}$.

\textit{Proof:} The proof of Theorem 1 is given in Appendix A. $ \hfill \blacksquare$

Since it is assumed that each receiver has only information about the channel correlation matrices and the statistical CSI, the conventional singular value decomposition (SVD) of channel matrices cannot be applied to MIMO data decoding. Thus, in this paper only the received signal at the $k$-th receive antenna of the relay is used to calculate the received SINR of symbol ${s_k}$, and successive interference cancellation (SIC) cannot be applied to data decoding. Therefore the interference to symbol $s_k$ cannot be cancelled and the received SINR for symbol $s_k$ at the relay is given by
\begin{equation} \label{eq:7}
{\gamma _{sr,k}} = \frac{{\frac{{G{p_{s,k}}}}{{r_{sr}^\alpha }}{{\left| {{\mathbf{h}}_{sr,k}^H{{\mathbf{g}}_{sr,k}}} \right|}^2}}}{{\sum\limits_{j = 1, j \ne k}^K {\frac{{G{p_{s,j}}}}{{r_{sr}^\alpha }}{{\left| {{\mathbf{h}}_{sr,k}^H{{\mathbf{g}}_{sr,j}}} \right|}^2} + {N_0}} }}
\end{equation}

In the second time slot, the BS keeps silent, the relay can either amplify or decode the received signal and forward it to the users.

\subsubsection{Amplify-and-Forward}
For AF relaying, in the second time slot the transmitted signal vector at the relay is
\begin{equation} \label{eq:8}
{{\mathbf{x}}_r} \!=\! \sum\limits_{m = 1}^K {\!\!\sqrt {{p_{r,m}}} {{\mathbf{g}}_{rd,m}}\!\left( {{\mathbf{h}}_{sr,m}^H\!\left( {\sum\limits_{n = 1}^K {\sqrt {\frac{{G{p_{s,n}}}}{{r_{sr}^\alpha }}} {{\mathbf{g}}_{sr,n}}{s_n}} } \!\right) \!+ {n_{r,m}}} \!\right)},
\end{equation}
and the received signal vector at the users is given by
\begin{eqnarray} \label{eq:9}
  && \!\!\!\!\!\!\!\!\!\!\!\!\!\! {{\mathbf{y}}_{rd}} \nonumber \\
  && \!\!\!\!\!\!\!\!\!\!\!\!\!\! = {{\mathbf{H}}_{rd}}{{\mathbf{x}}_r} + {{\mathbf{n}}_d} \nonumber \\
  && \!\!\!\!\!\!\!\!\!\!\!\!\!\! = \!\!\left[\!\!\! {\begin{array}{*{20}{c}}
  {{\mathbf{h}}_{rd,1}^H} \\
   \vdots  \\
  {{\mathbf{h}}_{rd,K}^H}
\end{array}} \!\!\!\!\right]\!\!\!\left[ {\sum\limits_{m = 1}^K {\!\!\sqrt {\frac{{G{p_{r,m}}}}{{r_{rd}^\alpha }}} {{\mathbf{g}}_{rd,m}}\!\!\left(\! {{\mathbf{h}}_{sr,m}^H\!\!\left( {\sum\limits_{n = 1}^K {\!\!\sqrt {\!\frac{{G{p_{s,n}}}}{{r_{sr}^\alpha }}} {{\mathbf{g}}_{sr,n}}{s_n}} } \!\!\right) \!\!+\! {n_{r,m}}} \!\!\right)} } \!\right] \nonumber \\
&&\!\!\!\!\!\!\!\!\!\! + \left[\!\! {\begin{array}{*{20}{c}}
  {{n_{d,1}}} \\
   \vdots  \\
  {{n_{d,K}}}
\end{array}} \!\!\right]
\end{eqnarray}
Thus in the second time slot the received signal at user $k$ is
\begin{eqnarray} \label{eq:10}
&& \!\!\!\!\!\!\!\!\!\!\!\!\!\!  {y_{rd,k}} \nonumber \\
&& \!\!\!\!\!\!\!\!\!\!\!\!\!\!=\! {\mathbf{h}}_{rd,k}^H\!\!\left( {\sum\limits_{m = 1}^K {\!\sqrt {\frac{{G{p_{r,m}}}}{{r_{rd}^\alpha }}} {{\mathbf{g}}_{rd,m}}\!\!\left(\! {{\mathbf{h}}_{sr,m}^H\!\!\left( {\sum\limits_{n = 1}^K {\!\sqrt {\frac{{G{p_{s,n}}}}{{r_{sr}^\alpha }}} {{\mathbf{g}}_{sr,n}}{s_n}} } \!\!\right) \!\!+\! {n_{r,m}}} \!\!\right)} } \!\right) \nonumber \\
&& \!\!\!\!\!\!\!\!\! +\, {n_{d,k}} \hfill \nonumber \\
&& \!\!\!\!\!\!\!\!\!\!\!\!\!\! = {\mathbf{h}}_{rd,k}^H\!\left( {\!\sqrt {\frac{{G{p_{r,1}}}}{{r_{rd}^\alpha }}} {{\mathbf{g}}_{rd,1}}{\mathbf{h}}_{sr,1}^H\!\left( {\sum\limits_{n = 1}^K {\!\sqrt {\frac{{G{p_{s,n}}}}{{r_{sr}^\alpha }}} {{\mathbf{g}}_{sr,n}}{s_n}} } \!\!\right)} \right. \hfill \nonumber \\
&& \!\!\!\!\!\!\!\!\!  +\!  \cdots  \!+\! \sqrt {\frac{{G{p_{r,K}}}}{{r_{rd}^\alpha }}} {{\mathbf{g}}_{rd,K}}{\mathbf{h}}_{sr,K}^H\!\left( {\sum\limits_{n = 1}^K {\!\sqrt {\frac{{G{p_{s,n}}}}{{r_{sr}^\alpha }}} {{\mathbf{g}}_{sr,n}}{s_n}} }\!\! \right)  \hfill \nonumber \\
&& \!\!\!\!\!\!\!\!\!  \left. {+ \sqrt {\frac{{G{p_{r,1}}}}{{r_{rd}^\alpha }}} {{\mathbf{g}}_{rd,1}}{n_{r,1}} +  \cdots  + \sqrt {\frac{{G{p_{r,K}}}}{{r_{rd}^\alpha }}} {{\mathbf{g}}_{rd,K}}{n_{r,K}}} \right) + {n_{d,k}}.
\end{eqnarray}
Then within two time slots the received SINR at user $k$ is expressed as Eq. (11),
\begin{figure*}[!t]
\normalsize \setlength{\arraycolsep}{0.5em}
\begin{eqnarray} \label{eq:11}
&&\!\!\!\!\!\!\!\!\!  \gamma _k^{AF} = \gamma _{rd,k}^{AF} \nonumber \\
&& \,\,\,   = \frac{{{{\left| {{\mathbf{h}}_{rd,k}^H\left( {\sum\limits_{m = 1}^K {\sqrt {\frac{{G{p_{r,m}}}}{{r_{rd}^\alpha }}} {{\mathbf{g}}_{rd,m}}{\mathbf{h}}_{sr,m}^H\sqrt {\frac{{G{p_{s,k}}}}{{r_{sr}^\alpha }}} {{\mathbf{g}}_{sr,k}}} } \right)} \right|}^2}}}{{{{\left| {{\mathbf{h}}_{rd,k}^H\!\!\left( {\sum\limits_{m = 1}^K {\sqrt {\frac{{G{p_{r,m}}}}{{r_{rd}^\alpha }}} {{\mathbf{g}}_{rd,m}}{\mathbf{h}}_{sr,m}^H\sum\limits_{j = 1, j \ne k}^K {\sqrt {\frac{{G{p_{s,j}}}}{{r_{sr}^\alpha }}} {{\mathbf{g}}_{sr,j}}} } } \right)} \right|}^2}\!\! +\! {{\left| {{\mathbf{h}}_{rd,k}^H\!\!\left( {\sum\limits_{m = 1}^K {\sqrt {\frac{{G{p_{r,m}}}}{{r_{rd}^\alpha }}} {{\mathbf{g}}_{rd,m}}} } \right)} \right|}^2}  {N_0} \! + \! {N_0}}} \nonumber \\
&&\!\!\!\!\!\!\!\!  \xrightarrow{{M\! \to\! \infty }}\!\!\frac{{\frac{{G{p_{s,k}}}}{{r_{sr}^\alpha }} \times \frac{{G{p_{r,k}}}}{{r_{rd}^\alpha }}{{\left| {{\mathbf{h}}_{rd,k}^H{{\mathbf{g}}_{rd,k}}} \right|}^2}{{\left| {{\mathbf{h}}_{sr,k}^H{{\mathbf{g}}_{sr,k}}} \right|}^2}}}{{\!\left(\!\! {\frac{{G{p_{r,k}}}}{{r_{rd}^\alpha }}{{\!\left| {{\mathbf{h}}_{rd,k}^H{{\mathbf{g}}_{rd,k}}} \right|}^2}\!\!\!\!\sum\limits_{j = 1,j \ne k}^K {\!\!\!\frac{{G{p_{s,j}}}}{{r_{sr}^\alpha }}{{\!\left| {{\mathbf{h}}_{sr,k}^H{{\mathbf{g}}_{sr,j}}} \right|}^2}} \!\! +\!\! {{\left| {{\mathbf{h}}_{rd,k}^H\!\!\!\sum\limits_{j = 1,j \ne k}^K {\!\!\!\sqrt {\frac{{G{p_{r,j}}}}{{r_{rd}^\alpha }}} {{\mathbf{g}}_{rd,j}}{\mathbf{h}}_{sr,j}^H\!\sqrt {\frac{{G{p_{s,j}}}}{{r_{sr}^\alpha }}} {{\mathbf{g}}_{sr,j}}} } \right|}^2}} \right) \!+\! \frac{{G{p_{r,k}}}}{{r_{rd}^\alpha }}{{\left| {{\mathbf{h}}_{rd,k}^H{{\mathbf{g}}_{rd,k}}} \right|}^2} \!{N_0} \!+\! {N_0}}} \nonumber \\
\end{eqnarray}
\setlength{\arraycolsep}{5pt}\hrulefill
\end{figure*}
and the average transmission rate for user $k$ is restricted by
\begin{equation} \label{eq:12}
\overline R _k^{AF} \le \frac{1}{2}\log \left( {1 + \gamma _k^{AF}} \right).
\end{equation}

\subsubsection{Decode-and-Forward}
In the second time slot, the BS keeps silent, the relay decodes and forwards the received signal to the users. The transmitted signal vector at the relay is
\begin{equation} \label{eq:13}
{{\mathbf{x}}_r} = \sum\limits_{n = 1}^K {\sqrt {{p_{r,n}}} {{\mathbf{g}}_{rd,n}}{s_n}},
\end{equation}
and the received signal vector at the users is given by
\begin{eqnarray} \label{eq:14}
  {{\mathbf{y}}_{rd}} &\!\!\!\!=\!\!\!\!& {{\mathbf{H}}_{rd}}{{\mathbf{x}}_r} + {{\mathbf{n}}_d} \nonumber \\
   &\!\!\!\!=\!\!\!\!& \left[\!\! {\begin{array}{*{20}{c}}
  {{\mathbf{h}}_{rd,1}^H} \\
   \vdots  \\
  {{\mathbf{h}}_{rd,K}^H}
\end{array}} \!\!\right]\left[ {\sum\limits_{n = 1}^K {\sqrt {\frac{{G{p_{r,n}}}}{{r_{rd}^\alpha }}} {{\mathbf{g}}_{rd,n}}{s_n}} } \right] + \left[\!\! {\begin{array}{*{20}{c}}
  {{n_{d,1}}} \\
   \vdots  \\
  {{n_{d,K}}}
\end{array}} \!\!\right] \nonumber \\
   &\!\!\!\!=\!\!\!\!& \left[\!\! {\begin{array}{*{20}{c}}
  {{\mathbf{h}}_{rd,1}^H\left( {\sum\limits_{n = 1}^K {\sqrt {\frac{{G{p_{r,n}}}}{{r_{rd}^\alpha }}} {{\mathbf{g}}_{rd,n}}{s_n}} } \right)} \\
   \vdots  \\
  {{\mathbf{h}}_{rd,K}^H\left( {\sum\limits_{n = 1}^K {\sqrt {\frac{{G{p_{r,n}}}}{{r_{rd}^\alpha }}} {{\mathbf{g}}_{rd,n}}{s_n}} } \right)}
\end{array}} \!\!\right] + \left[\!\! {\begin{array}{*{20}{c}}
  {{n_{d,1}}} \\
   \vdots  \\
  {{n_{d,K}}}
\end{array}} \!\!\right]
\end{eqnarray}
Thus in the second time slot the received signal at user $k$ is
\begin{equation} \label{eq:15}
{y_{rd,k}} = {\mathbf{h}}_{rd,k}^H\left( {\sum\limits_{n = 1}^K {\sqrt {\frac{{G{p_{r,n}}}}{{r_{rd}^\alpha }}} {{\mathbf{g}}_{rd,n}}{s_n}} } \right) + {n_{d,k}},
\end{equation}
and the received SINR at user $k$ is given by
\begin{equation} \label{eq:16}
\gamma _{rd,k}^{DF} = \frac{{\frac{{G{p_{r,k}}}}{{r_{rd}^\alpha }}{{\left| {{\mathbf{h}}_{rd,k}^H{{\mathbf{g}}_{rd,k}}} \right|}^2}}}{{\sum\limits_{n = 1,n \ne k}^K {\frac{{G{p_{r,n}}}}{{r_{rd}^\alpha }}{{\left| {{\mathbf{h}}_{rd,k}^H{{\mathbf{g}}_{rd,n}}} \right|}^2} + {N_0}} }}
\end{equation}
Then within two time slots the received SINR at user $k$ is expressed as
\begin{eqnarray} \label{eq:17}
  && \!\!\!\!\!\!\!\!\!\!\!\!\!\!\!\!\!\!\! \gamma _k^{DF} = \min \left\{ {\gamma _{sr,k}^{DF},{\text{ }}\gamma _{rd,k}^{DF}} \right\} \nonumber \\
  && \!\!\!\!\!\! = \min \left\{ {\frac{{\frac{{G{p_{s,k}}}}{{r_{sr}^\alpha }}{{\left| {{\mathbf{h}}_{sr,k}^H{{\mathbf{g}}_{sr,k}}} \right|}^2}}}{{\sum\limits_{n = 1,n \ne k}^K {\frac{{G{p_{s,n}}}}{{r_{sr}^\alpha }}{{\left| {{\mathbf{h}}_{sr,k}^H{{\mathbf{g}}_{sr,n}}} \right|}^2} + {N_0}} }},} \right. \hfill \nonumber \\
  && \quad\qquad  \left.{ \frac{{\frac{{G{p_{r,k}}}}{{r_{rd}^\alpha }}{{\left| {{\mathbf{h}}_{rd,k}^H{{\mathbf{g}}_{rd,k}}} \right|}^2}}}{{\sum\limits_{n = 1,n \ne k}^K {\frac{{G{p_{r,n}}}}{{r_{rd}^\alpha }}{{\left| {{\mathbf{h}}_{rd,k}^H{{\mathbf{g}}_{rd,n}}} \right|}^2}}  + {N_0}}}} \right\}
\end{eqnarray}
Within two time slots the average transmission rate for user $k$ is restricted by
\begin{equation} \label{eq:18}
\overline R _k^{DF} \le \frac{1}{2}\log \left( {1 + \gamma _k^{DF}} \right)
\end{equation}

\section{Deterministic Equivalent for the SINR}
This section discusses the deterministic approximation for the SINR at the users. Let ${\mathbf{\Phi }} = \left\{ {{{\mathbf{h}}_{sr,1}}, \cdots ,{{\mathbf{h}}_{sr,K}},{{\mathbf{h}}_{rd,1}}, \cdots ,{{\mathbf{h}}_{rd,K}}} \right\}$, ${\mathbf{\hat \Phi }} = \left\{ {{{{\mathbf{\hat h}}}_{sr,1}}, \! \cdots \! ,{{{\mathbf{\hat h}}}_{sr,K}},} \right. \left. {{{{\mathbf{\hat h}}}_{rd,1}}, \cdots ,{{{\mathbf{\hat h}}}_{rd,K}}} \right\}$, ${{\mathbf{h}}_k} \in {\mathbf{\Phi }}$, ${{\mathbf{\hat h}}_k} \in {\mathbf{\hat \Phi }}$, before we derive the deterministic equivalent for the SINR at user $k$ in M-MIMO-ADF relaying with RZF precoding, let's first derive the deterministic equivalents for the following items:

\subsection{Deterministic Equivalent of ${\mathbf{h}}_k^H{{\mathbf{\hat W}}_l}{{\mathbf{\hat h}}_k}$, ${\mathbf{\hat h}}_k^H{\mathbf{\hat W}}_l^2{{\mathbf{\hat h}}_k}$ and $\sum\limits_{j = 1, j \ne k}^K {{{\left| {{\mathbf{h}}_k^H{{{\mathbf{\hat W}}}_l}{{{\mathbf{\hat h}}}_j}} \right|}^2}}$}
On how to obtain the deterministic equivalent of ${\mathbf{h}}_k^H{{\mathbf{\hat W}}_l}{{\mathbf{\hat h}}_k}$, ${\mathbf{\hat h}}_k^H{\mathbf{\hat W}}_l^2{{\mathbf{\hat h}}_k}$ and $\sum\limits_{j = 1,j \ne k}^K {{{\left| {{\mathbf{h}}_k^H{{{\mathbf{\hat W}}}_l}{{{\mathbf{\hat h}}}_j}} \right|}^2}} $ for $\forall k \in \left\{ {1, \cdots ,K} \right\}$, $l \in \left\{ {1,2} \right\}$, and $M \gg K$, let's first recall the results of [21]:
\begin{equation} \label{eq:19}
{\mathbf{h}}_k^H{{\mathbf{\hat W}}_l}{{\mathbf{\hat h}}_k} - \sqrt {1 - \tau _k^2} \frac{{m_k^o}}{{1 + m_k^o}}\xrightarrow{{M \to \infty }}0,
\end{equation}
\begin{equation} \label{eq:20}
{\mathbf{\hat h}}_k^H{\mathbf{\hat W}}_l^2{{\mathbf{\hat h}}_k} - \frac{{{{m'}_k}}}{{M{{\left( {1 + m_k^o} \right)}^2}}}\xrightarrow{{M \to \infty }}0,
\end{equation}
\begin{equation} \label{eq:21}
\sum\limits_{j = 1,j \ne k}^K {{{\left| {{\mathbf{h}}_k^H{{{\mathbf{\hat W}}}_l}{{{\mathbf{\hat h}}}_j}} \right|}^2}}  - \Upsilon _k^o{\Phi _k}\xrightarrow{{M \to \infty }}0,
\end{equation}
almost surely, where
\begin{equation} \label{eq:22}
m_k^o = \frac{1}{M}{\text{tr}}\left( {{{\left( {{\mathbf{\Theta }}_k^{{1 \mathord{\left/
 {\vphantom {1 2}} \right.
 \kern-\nulldelimiterspace} 2}}} \right)}^H}{\mathbf{\Theta }}_k^{{1 \mathord{\left/
 {\vphantom {1 2}} \right.
 \kern-\nulldelimiterspace} 2}}} \right){\mathbf{T}},
\end{equation}
\begin{equation} \label{eq:23}
{\mathbf{T}} = {\left( {\frac{1}{M}\sum\limits_{j = 1}^K {\frac{{{{\left( {{\mathbf{\Theta }}_j^{{1 \mathord{\left/
 {\vphantom {1 2}} \right.
 \kern-\nulldelimiterspace} 2}}} \right)}^H}{\mathbf{\Theta }}_j^{{1 \mathord{\left/
 {\vphantom {1 2}} \right.
 \kern-\nulldelimiterspace} 2}}}}{{1 + m_j^o}}}  + {\alpha _l}{{\mathbf{I}}_M}} \right)^{ - 1}},
\end{equation}
\begin{equation} \label{eq:24}
\Upsilon _k^o = \frac{1}{M}\sum\limits_{j = 1,j \ne k}^K {\frac{{{{m'}_{j,k}}}}{{{{\left( {1 + m_j^o} \right)}^2}}}},
\end{equation}
\begin{equation} \label{eq:25}
{\Phi _k} = \frac{{1 - \tau _k^2\left( {1 - {{\left( {1 + m_k^o} \right)}^2}} \right)}}{{{{\left( {1 + m_k^o} \right)}^2}}},
\end{equation}
with ${\mathbf{m'}} = {\left[ {{{m'}_1}, \cdots ,{{m'}_K}} \right]^T}$ and ${{\mathbf{m'}}_k} = {\left[ {{{m'}_{1,k}}, \cdots ,{{m'}_{K,k}}} \right]^T}$ defined by
\begin{equation} \label{eq:26}
{\mathbf{m'}} = {\left( {{{\mathbf{I}}_K} - {\mathbf{J}}} \right)^{ - 1}}{\mathbf{v}},\quad\quad {{\mathbf{m'}}_k} = {\left( {{{\mathbf{I}}_K} - {\mathbf{J}}} \right)^{ - 1}}{{\mathbf{v}}_k},
\end{equation}
where $\mathbf{J}$, $\mathbf{v}$ and ${{\mathbf{v}}_k}$ are given by
\begin{equation} \label{eq:27}
{\left[ {\mathbf{J}} \right]_{ij}} = \frac{{\frac{1}{M}{\text{tr}}\left( {{{\left( {{\mathbf{\Theta }}_i^{{1 \mathord{\left/
 {\vphantom {1 2}} \right.
 \kern-\nulldelimiterspace} 2}}} \right)}^H}{\mathbf{\Theta }}_i^{{1 \mathord{\left/
 {\vphantom {1 2}} \right.
 \kern-\nulldelimiterspace} 2}}} \right){\mathbf{T}}\left( {{{\left( {{\mathbf{\Theta }}_j^{{1 \mathord{\left/
 {\vphantom {1 2}} \right.
 \kern-\nulldelimiterspace} 2}}} \right)}^H}{\mathbf{\Theta }}_j^{{1 \mathord{\left/
 {\vphantom {1 2}} \right.
 \kern-\nulldelimiterspace} 2}}} \right){\mathbf{T}}}}{{M{{\left( {1 + m_j^o} \right)}^2}}},
\end{equation}
\begin{equation} \label{eq:28}
{\mathbf{v}} \!=\!\! {\left[\!\!\! {\begin{array}{*{20}{c}}
  {\frac{1}{M}{\text{tr}}\!\left( {{{\left( {{\mathbf{\Theta }}_1^{{1 \mathord{\left/
 {\vphantom {1 2}} \right.
 \kern-\nulldelimiterspace} 2}}} \right)}^{\!H}}\!\!{\mathbf{\Theta }}_1^{{1 \mathord{\left/
 {\vphantom {1 2}} \right.
 \kern-\nulldelimiterspace} 2}}} \!\right)\!{{\mathbf{T}}^2},}{ \,\cdots ,}{\frac{1}{M}{\text{tr}}\!\left( {{{\left( {{\mathbf{\Theta }}_K^{{1 \mathord{\left/
 {\vphantom {1 2}} \right.
 \kern-\nulldelimiterspace} 2}}} \!\right)}^{\!H}}\!\!{\mathbf{\Theta }}_K^{{1 \mathord{\left/
 {\vphantom {1 2}} \right.
 \kern-\nulldelimiterspace} 2}}} \right)\!{{\mathbf{T}}^2}}
\end{array}} \!\!\!\right]^{\!T}}\!\!,
\end{equation}
\begin{eqnarray} \label{eq:29}
 && \!\!\!\!\!\!\!\! {{\mathbf{v}}_k} = \left[ {\frac{1}{M}{\text{tr}}\left( {{\mathbf{\Theta }}_1^{{1 \mathord{\left/
 {\vphantom {1 2}} \right.
 \kern-\nulldelimiterspace} 2}}{\mathbf{\Theta }}_1^{{1 \mathord{\left/
 {\vphantom {1 2}} \right.
 \kern-\nulldelimiterspace} 2}}} \right){\mathbf{T}}\left( {{\mathbf{\Theta }}_k^{{1 \mathord{\left/
 {\vphantom {1 2}} \right.
 \kern-\nulldelimiterspace} 2}}{\mathbf{\Theta }}_k^{{1 \mathord{\left/
 {\vphantom {1 2}} \right.
 \kern-\nulldelimiterspace} 2}}} \right){\mathbf{T}},} \right. \hfill \nonumber \\
 && \quad {\left. { \cdots ,{\text{ }}\frac{1}{M}{\text{tr}}\left( {{\mathbf{\Theta }}_K^{{1 \mathord{\left/
 {\vphantom {1 2}} \right.
 \kern-\nulldelimiterspace} 2}}{\mathbf{\Theta }}_K^{{1 \mathord{\left/
 {\vphantom {1 2}} \right.
 \kern-\nulldelimiterspace} 2}}} \right){\mathbf{T}}\left( {{\mathbf{\Theta }}_k^{{1 \mathord{\left/
 {\vphantom {1 2}} \right.
 \kern-\nulldelimiterspace} 2}}{\mathbf{\Theta }}_k^{{1 \mathord{\left/
 {\vphantom {1 2}} \right.
 \kern-\nulldelimiterspace} 2}}} \right){\mathbf{T}}} \right]^T} \hfill
\end{eqnarray}

Different from the iterative method in [21] to obtain the deterministic equivalent of ${\mathbf{h}}_k^H{{\mathbf{\hat W}}_l}{{\mathbf{\hat h}}_k}$, which may be time-consuming and may not function well for a real massive MIMO system, in this paper we obtain a closed-form expression for the deterministic equivalent of ${\mathbf{h}}_k^H{{\mathbf{\hat W}}_l}{{\mathbf{\hat h}}_k}$. It requires less computation time and is more practicable for a real massive MIMO system which may be sensitive to delay.

The following theorem and proposition are useful for the subsequent large system analysis.

\textit{\!Theorem 2:} Let ${\mathbf{\hat \Phi }} = \left\{ {{{{\mathbf{\hat h}}}_{sr,1}}, \cdots ,{{{\mathbf{\hat h}}}_{sr,K}},{{{\mathbf{\hat h}}}_{rd,1}}, \cdots ,{{{\mathbf{\hat h}}}_{rd,K}}} \right\}$, ${{\mathbf{\hat h}}_k} \in {\mathbf{\hat \Phi }}$, ${{\mathbf{\hat h}}_{k'}} \in {\mathbf{\hat \Phi }} - {{\mathbf{\hat h}}_k}$, then we have $\left| {{{{\mathbf{\hat h}}}_k^H}{\mathbf{\hat W}}_l^2{{{\mathbf{\hat h}}}_k}} \right| \gg \left| {{{{\mathbf{\hat h}}}_k^H}{\mathbf{\hat W}}_l^2{{{\mathbf{\hat h}}}_{k'}}} \right|$ for any $k \ne k'$, $M \to \infty $ and $M \gg K$, where $k,k' \in \left\{ {1, \cdots ,K} \right\}$ and $l \in \left\{ {1,2} \right\}$.

\textit{\!Proof:} The proof of Theorem 2 is given in Appendix B. $ \hfill \blacksquare$

\textit{\!Proposition 1:} Let ${\mathbf{\Phi }} \!=\! \left\{ {{{\mathbf{h}}_{sr,1}},\! \cdots \!,{{\mathbf{h}}_{sr,K}},{{\mathbf{h}}_{rd,1}}, \! \cdots \! ,{{\mathbf{h}}_{rd,K}}} \right\}$, ${\mathbf{\hat \Phi }} \!=\! \left\{ {{{{\mathbf{\hat h}}}_{sr,1}}, \! \cdots \! ,{{{\mathbf{\hat h}}}_{sr,K}},{{{\mathbf{\hat h}}}_{rd,1}}, \! \cdots \! ,{{{\mathbf{\hat h}}}_{rd,K}}} \right\}$, ${{\mathbf{h}}_k} \in {\mathbf{\Phi }}$, ${{\mathbf{\hat h}}_k} \in {\mathbf{\hat \Phi }}$, then we have ${\mathbf{h}}_k^H{{\mathbf{\hat W}}_l}{{\mathbf{\hat h}}_k} - \frac{{\sqrt {1 - \tau _k^2} {\text{tr}}\left( {{{\left( {{\mathbf{\Theta }}_k^{{1 \mathord{\left/
 {\vphantom {1 2}} \right.
 \kern-\nulldelimiterspace} 2}}} \right)}^H}{\mathbf{\Theta }}_k^{{1 \mathord{\left/
 {\vphantom {1 2}} \right.
 \kern-\nulldelimiterspace} 2}}} \right)}}{{M{\alpha _l} + {\text{tr}}\left( {{{\left( {{\mathbf{\Theta }}_k^{{1 \mathord{\left/
 {\vphantom {1 2}} \right.
 \kern-\nulldelimiterspace} 2}}} \right)}^H}{\mathbf{\Theta }}_k^{{1 \mathord{\left/
 {\vphantom {1 2}} \right.
 \kern-\nulldelimiterspace} 2}}} \right)}}\xrightarrow{{M \to \infty }}0$ for any $M, K \to \infty $ and $M \gg K$, where $k \in \left\{ {1, \cdots ,K} \right\}$ and $l \in \left\{ {1,2} \right\}$.

\textit{\!Proof:}The proof of Proposition 1 is given in Appendix C. $ \hfill\!\blacksquare$

\subsection{Transmit Power}
\subsubsection{Transmit Power Constraint at the BS}
To satisfy the total transmit power constraint at the BS, the precoding vectors are normalized as
\begin{equation} \label{eq:30}
\sum\limits_{k = 1}^K {{p_{s,k}}{\mathbf{g}}_{sr,k}^H{{\mathbf{g}}_{sr,k}}}  = \sum\limits_{k = 1}^K {{p_{s,k}}\xi _1^2{\mathbf{\hat h}}_{sr,k}^H{\mathbf{\hat W}}_1^2{{{\mathbf{\hat h}}}_{sr,k}}}  \le P
\end{equation}
Then we have the following constraint for $\xi _1^2$:
\begin{equation} \label{eq:31}
\xi _1^2 - {\left( {\xi _1^o} \right)^2}\xrightarrow{{M \to \infty }}0
\end{equation}
almost surely, where
\begin{equation} \label{eq:32}
{\left( {\xi _1^o} \right)^2} = \frac{P}{{\sum\limits_{k = 1}^K {{p_{s,k}}\left( {\frac{{{{m'}_{sr,k}}}}{{M{{\left( {1 + m_{sr,k}^o} \right)}^2}}}} \right)} }}
\end{equation}

\subsubsection{Transmit Power Constraint at the Relay for AF Relaying}
To satisfy the total transmit power constraint for AF relaying, the precoding vectors for the transmitted signals at the relay are normalized as
\begin{eqnarray} \label{eq:33}
&& \!\!\!\!\!\!\!\!\!\!\!\!\! \sum\limits_{m = 1}^K {{p_{r,m}}{\mathbf{g}}_{rd,m}^H{{\mathbf{g}}_{rd,m}}\!\!\left( {\sum\limits_{n = 1}^K {{{\left| {{\mathbf{h}}_{sr,m}^H\frac{{G{p_{s,n}}}}{{r_{sr}^\alpha }}{{\mathbf{g}}_{sr,n}}} \right|}^2}}  + {N_0}} \right)} \!\!  \xrightarrow[{\left( a \right)}]{{M \to \infty }} \nonumber \\
&& \!\!\!\!\!\!\!\!\!\!\!\!\! \xi _{AF}^2\!\!\sum\limits_{m = 1}^K \!\!{{p_{r,m}}\!\!\left( {{\mathbf{\hat h}}_{rd,m}^H{\!\!\mathbf{\hat W}}_2^2{{{\mathbf{\hat h}}}_{rd,m}}} \!\right)\!\!\left(\! {\frac{{G{p_{s,m}}}}{{r_{sr}^\alpha }}\xi _1^2{{\left| {{\mathbf{h}}_{sr,m}^H{{{\!\!\mathbf{\hat W}}}_1}{{{\mathbf{\hat h}}}_{sr,m}}} \right|}^2} \!\!\!+ \!\! {N_0}} \!\right)}  \nonumber \\
&& \!\!\!\!\!\!\!\!\!\!\!\! \le P
\end{eqnarray}
where (a) follows Theorem 1. Then we have the following constraint for $\xi _{AF}^2$:
\begin{equation} \label{eq:34}
\xi _{AF}^2 - {\left( {\xi _{AF}^o} \right)^2}\xrightarrow{{M \to \infty }}0
\end{equation}
almost surely, where
\begin{equation} \label{eq:35}
{\left( {\xi _{AF}^o} \right)^{\!2}} \!=\! \frac{P}{{\!\sum\limits_{m = 1}^K { \!\!\frac{{p_{r,m}}{{{m'}_{rd,m}}}}{{M{{\left( {1 + m_{rd,m}^o} \right)}^2}}}\!\!\left[\! {\frac{{G{p_{s,m}}}}{{r_{sr}^\alpha }}{{\!\left( {\xi _1^o} \right)}^{\!2}}\!\!\left( {1 - \tau _m^2} \right)\!\!{{\left(\! {\frac{{m_{sr,m}^o}}{{1 + m_{sr,m}^o}}} \!\right)}^{\!2}}\!\!\!\! +\!\! {N_0}} \!\right]} }}
\end{equation}

\subsubsection{Transmit Power Constraint at the Relay for DF Relaying}
To satisfy the total transmit power constraint for DF relaying, the precoding vectors for the transmitted signals at the relay are normalized as
\begin{equation} \label{eq:36}
\sum\limits_{n = 1}^K {{p_{r,n}}{\mathbf{g}}_{rd,n}^H{{\mathbf{g}}_{rd,n}}}  = \xi _{DF}^2\sum\limits_{n = 1}^K {{p_{r,n}}\left( {{\mathbf{\hat h}}_{rd,n}^H{\mathbf{\hat W}}_2^2{{{\mathbf{\hat h}}}_{rd,n}}} \right) \le P}
\end{equation}
Then we have the following constraint for $\xi _{DF}^2$:
\begin{equation} \label{eq:37}
\xi _{DF}^2 - {\left( {\xi _{DF}^o} \right)^2}\xrightarrow{{M \to \infty }}0
\end{equation}
almost surely, where
\begin{equation} \label{eq:38}
{\left( {\xi _{DF}^o} \right)^2} = \frac{P}{{\sum\limits_{n = 1}^K {{p_{r,n}}\frac{{{{m'}_{rd,n}}}}{{M{{\left( {1 + m_{rd,n}^o} \right)}^2}}}} }}
\end{equation}

\subsection{Deterministic Equivalent for the SINR}
\textit{Corollary 1:} Let ${\mathbf{\Phi }} \!=\! \left\{ {{{\mathbf{h}}_{sr,1}},\! \cdots \!,{{\mathbf{h}}_{sr,K}},{{\mathbf{h}}_{rd,1}}, \! \cdots \! ,{{\mathbf{h}}_{rd,K}}} \right\}$, ${\mathbf{\hat \Phi }} \!=\! \left\{ {{{{\mathbf{\hat h}}}_{sr,1}}, \! \cdots \! ,{{{\mathbf{\hat h}}}_{sr,K}},{{{\mathbf{\hat h}}}_{rd,1}}, \! \cdots \! ,{{{\mathbf{\hat h}}}_{rd,K}}} \right\}$, ${{\mathbf{h}}_k} \in {\mathbf{\Phi }}$, ${{\mathbf{\hat h}}_k} \in {\mathbf{\hat \Phi }}$ for any $k \ne k'$, $M \to \infty $ and $M \gg K$ where $k,k' \in \left\{ {1, \cdots ,K} \right\}$ and $l \in \left\{ {1,2} \right\}$, if the transmit power for each user message at the BS or the relay is equally allocated as ${P \mathord{\left/
 {\vphantom {P K}} \right.
 \kern-\nulldelimiterspace} K}$, then the deterministic equivalent for the received SINR at user $k$ in M-MIMO-ADF relaying is given by
\begin{equation} \label{eq:39}
\gamma _k^{AF} - {\left( {\gamma _k^{AF}} \right)^o}\xrightarrow{{M \to \infty }}0
\end{equation}
\begin{equation} \label{eq:40}
\gamma _k^{DF} - {\left( {\gamma _k^{DF}} \right)^o}\xrightarrow{{M \to \infty }}0
\end{equation}
almost surely, where ${\left( {\gamma _k^{AF}} \right)^o}$ and ${\left( {\gamma _k^{DF}} \right)^o}$ are given by Eq. (41) and Inequality (42)
\begin{figure*}[!t]
\normalsize \setlength{\arraycolsep}{0.5em}
\begin{equation} \label{eq:41}
\!\!\!\!\! {\left(\! {\gamma _k^{AF}} \right)\!^o} \!\!= \!\!\frac{{\frac{G}{{r_{sr}^\alpha }}\frac{G}{{r_{rd}^\alpha }}{{\left( {\frac{P}{K}} \right)}^2}{{\left( {\xi _1^o} \right)}^2}{{\!\left( {\xi _{AF}^o} \!\right)}^2}\Gamma_{sr,k}^2\Gamma_{rd,k}^2}}{{\frac{G}{{r_{rd}^\alpha }}\frac{G}{{r_{sr}^\alpha }}{{\!\left(\! {\frac{P}{K}} \!\right)}^2}{{\!\left( {\xi _1^o} \!\right)}^2}{{\!\left( {\xi _{AF}^o} \!\right)}^2}\Gamma_{rd,k}^2\!\Upsilon _{sr,k}^o{\!\Phi _{sr,k}} \!\!+\!\! \frac{G}{{r_{sr}^\alpha }}\frac{G}{{r_{rd}^\alpha }}{{\!\left(\! {\frac{P}{K}} \!\right)}^2}{{\!\left( {\xi _1^o} \!\right)}^2}{{\!\left( {\xi _{AF}^o} \!\right)}^2}\Gamma_{sr,k}^2\!\Upsilon _{rd,k}^o{\!\Phi _{rd,k}} \!\!+\!\! \frac{G}{{r_{rd}^\alpha }}\frac{P}{K}{{\!\left( {\xi _{AF}^o} \!\right)}^2}\Gamma_{rd,k}^2{N_0} \!\!+\!\! {N_0}}} \\
\end{equation}
\begin{equation} \label{eq:42}
{\left( {\gamma _k^{DF}} \right)^o} \le \min \left\{ {\frac{{\frac{G}{{r_{sr}^\alpha }}\frac{P}{K}{{\left( {\xi _1^o} \right)}^2}\Gamma_{sr,k}^2}}{{\frac{G}{{r_{sr}^\alpha }}\frac{P}{K}{{\left( {\xi _1^o} \right)}^2}\Upsilon _{sr,k}^o{\Phi _{sr,k}} + {N_0}}},{\text{ }}\frac{{\frac{G}{{r_{rd}^\alpha }}\frac{P}{K}{{\left( {\xi _{DF}^o} \right)}^2}\Gamma_{rd,k}^2}}{{\frac{G}{{r_{rd}^\alpha }}\frac{P}{K}{{\left( {\xi _{DF}^o} \right)}^2}\Upsilon _{rd,k}^o{\Phi _{rd,k}} + {N_0}}}} \right\}
\end{equation}
\setlength{\arraycolsep}{5pt}\hrulefill
\end{figure*}
with
\begin{eqnarray} \label{eq:43}
{\Gamma _{sr,k}} = \sqrt {1 - \tau _k^2} \frac{{{\text{tr}}{{\left( {{\mathbf{\Theta }}_k^{{1 \mathord{\left/
 {\vphantom {1 2}} \right.
 \kern-\nulldelimiterspace} 2}}} \right)}^H}{\mathbf{\Theta }}_k^{{1 \mathord{\left/
 {\vphantom {1 2}} \right.
 \kern-\nulldelimiterspace} 2}}}}{{M{\alpha _1} + {\text{tr}}{{\left( {{\mathbf{\Theta }}_k^{{1 \mathord{\left/
 {\vphantom {1 2}} \right.
 \kern-\nulldelimiterspace} 2}}} \right)}^H}{\mathbf{\Theta }}_k^{{1 \mathord{\left/
 {\vphantom {1 2}} \right.
 \kern-\nulldelimiterspace} 2}}}}, \\
{\Gamma _{rd,k}} = \sqrt {1 - \tau _k^2} \frac{{{\text{tr}}{{\left( {{\mathbf{\Theta }}_k^{{1 \mathord{\left/
 {\vphantom {1 2}} \right.
 \kern-\nulldelimiterspace} 2}}} \right)}^H}{\mathbf{\Theta }}_k^{{1 \mathord{\left/
 {\vphantom {1 2}} \right.
 \kern-\nulldelimiterspace} 2}}}}{{M{\alpha _2} + {\text{tr}}{{\left( {{\mathbf{\Theta }}_k^{{1 \mathord{\left/
 {\vphantom {1 2}} \right.
 \kern-\nulldelimiterspace} 2}}} \right)}^H}{\mathbf{\Theta }}_k^{{1 \mathord{\left/
 {\vphantom {1 2}} \right.
 \kern-\nulldelimiterspace} 2}}}},
\end{eqnarray}
${\Upsilon _{sr,k}^o}$ and ${\Upsilon _{rd,k}^o}$ are defined as Eq. (24), ${{\Phi _{sr,k}}}$ and ${{\Phi _{rd,k}}}$ are defined as Eq. (25), ${{\left( {\xi_1^o} \right)}^2}$, ${{\left( {\xi_ {AF}^o} \right)}^2}$, ${{{\left( {\xi _{DF}^o} \right)}^2}}$ are defined in Eq. (32), Eq. (35), Eq. (38), respectively.

\textit{Proof:} Please note that the derived SINRs have to be averaged over to get the ergodic ones. Eq. (41) can be obtained from Eq. (11) and section III. (A)(B), and the two terms in the min\{\} operator of Inequality (42) can be obtained from Eq. (17) and section III. (A)(B). Next we complete the proof of Inequality (42) to show that the minimum one of the two terms in the min\{\} operator is a lower bound for ${\left( {\gamma _k^{DF}} \right)^o}$: Applying Theorem 1 we have $\left| {{{\mathbf{h}}_k^H}{{{\mathbf{\hat W}}}_l}{{{\mathbf{\hat h}}}_k}} \right| \gg \left| {{{\mathbf{h}}_k^H}{{{\mathbf{\hat W}}}_l}{{{\mathbf{\hat h}}}_{k'}}} \right|$ if the assumption holds true (for any $k \ne k'$, $M \to \infty $ and $M \gg K$ where $k,k' \in \left\{ {1, \cdots ,K} \right\}$ and $l \in \left\{ {1,2} \right\}$). From Eq. (49) and Eq. (50) of Appendix A, if the signal channel coefficient $ \uparrow \left| {{\mathbf{h}}_k^H{{{\mathbf{\hat W}}}_l}{{{\mathbf{\hat h}}}_k}} \right| \to t\left| {{\mathbf{h}}_k^H{{{\mathbf{\hat W}}}_l}{{{\mathbf{\hat h}}}_k}} \right|$ for $\forall t > 1$, then the interference channel coefficient $ \uparrow \left| {{\mathbf{h}}_k^H{{{\mathbf{\hat W}}}_l}{{{\mathbf{\hat h}}}_{k'}}} \right|$ will be far less than $ t\left| {{\mathbf{h}}_k^H{{{\mathbf{\hat W}}}_l}{{{\mathbf{\hat h}}}_{k'}}} \right|$, so that we have the following inequality:
\begin{eqnarray} \label{eq:45}
  \varphi \left( {X = \left| {{\mathbf{h}}_k^H{{{\mathbf{\hat W}}}_l}{{{\mathbf{\hat h}}}_k}} \right|} \right) &\!\!\!=\!\!\!& \frac{{\frac{{GP}}{K}{{\left| {{\mathbf{h}}_k^H{{{\mathbf{\hat W}}}_l}{{{\mathbf{\hat h}}}_k}} \right|}^2}}}{{\sum\limits_{n = 1,n \ne k}^K {\frac{{GP}}{K}{{\left| {{\mathbf{h}}_k^H{{{\mathbf{\hat W}}}_l}{{{\mathbf{\hat h}}}_n}} \right|}^2}}  + {N_0}}} \hfill \nonumber \\
    &\!\!\!\le \!\!\!& \frac{{\frac{{GP}}{K}t{{\left| {{\mathbf{h}}_k^H{{{\mathbf{\hat W}}}_l}{{{\mathbf{\hat h}}}_k}} \right|}^2}}}{{\sum\limits_{n = 1,n \ne k}^K {\frac{{GP}}{K}t{{\left| {{\mathbf{h}}_k^H{{{\mathbf{\hat W}}}_l}{{{\mathbf{\hat h}}}_n}} \right|}^2}}  + {N_0}}} \nonumber \\
    &\!\!\!=\!\!\!& \varphi \left( {X = \sqrt t \left| {{\mathbf{h}}_k^H{{{\mathbf{\hat W}}}_l}{{{\mathbf{\hat h}}}_k}} \right|} \right), \hfill
\end{eqnarray}
 which proves that $\varphi \left( X \right)$ is a monotone increasing quasiconcave function. Besides, from Eq. (49) and Eq. (50) of Appendix A, $ X =  \left| {{\mathbf{h}}_k^H{{{\mathbf{\hat W}}}_l}{{{\mathbf{\hat h}}}_k}} \right|$ can be seen as a monotone increasing concave function, while $ Y =  \left| {{\mathbf{h}}_k^H{{{\mathbf{\hat W}}}_l}{{{\mathbf{\hat h}}}_{k'}}} \right|$ can be seen as a constant if the assumption holds true, therefore it further proves that $\varphi \left( X \right)$ is a monotone increasing concave function. Since $\gamma _{sr,k}^{DF}\left( X \right)$ and $\gamma _{rd,k}^{DF}\left( X \right)$ has the same form as $\varphi \left( X \right)$ so that they are both concave functions if the assumption holds true, then for $\forall\theta  \in \left[ {0,1} \right]$ we have the following inequality:
\begin{eqnarray} \label{eq:46}
  &&\!\!\!\!\!\!\!\!\!\!\!\! \gamma _k^{DF}\left( {\theta{X_1} + \left( {1 - \theta} \right){X_2}} \right) \nonumber \\
  &&\!\!\!\!\!\!\!\!\!\!\!\!= \min \left\{ {\gamma _{sr,k}^{DF}\left( {\theta{X_1} + \left( {1 - \theta} \right){X_2}} \right),\,
  \gamma _{rd,k}^{DF}\left( {\theta{X_1} + \left( {1 - \theta} \right){X_2}} \right)} \right\} \nonumber \\
   &&\!\!\!\!\!\!\!\!\!\!\!\! \ge  \min \!\left\{ {\theta\gamma _{sr,k}^{DF}\left( {{X_1}} \right) + \left( {1 - \theta} \right)\gamma _{sr,k}^{DF}\left( {{X_2}} \right), }\right. \hfill \nonumber \\
   &&\quad\; \left. {\theta\gamma _{rd,k}^{DF}\left( {{X_1}} \right) + \left( {1 - \theta} \right)\gamma _{rd,k}^{DF}\left( {{X_2}} \right)} \right\} \nonumber \\
   &&\!\!\!\!\!\!\!\!\!\!\!\! \ge  \theta\min \! \left\{ {\gamma _{sr,k}^{DF}\left( {{X_1}} \right),{\text{ }}\gamma _{rd,k}^{DF}\left( {{X_1}} \right)} \right\}\nonumber \\
   &&\!\!\!\!\! + \left( {1 - \theta} \right)\min \! \left\{ {\gamma _{sr,k}^{DF}\left( {{X_2}} \right),{\text{ }}\gamma _{rd,k}^{DF}\left( {{X_2}} \right)} \right\} \nonumber \\
   &&\!\!\!\!\!\!\!\!\!\!\!\! = \theta\gamma _k^{DF}\left( {{X_1}} \right) + \left( {1 - \theta} \right)\gamma _k^{DF}\left( {{X_2}} \right), \hfill
\end{eqnarray}
which proves that $\gamma _k^{DF}\left( X \right)$ is a concave function if the assumption holds true. From Jensen's inequality and the concavity of $\gamma _k^{DF}\left( X \right)$ it leads to the inequality $E\left( {\gamma _k^{DF}\left( X \right)} \right) \le \gamma _k^{DF}\left( {E\left( X \right)} \right)$ if the assumption holds true, which completes the proof.$\hfill \blacksquare$

\section{Simulation Results}
In this section, Monte-Carlo (MC) simulation results are compared with large system approximations to validate the accuracy of the derived deterministic equivalents, and demonstrate the performance of the proposed massive MIMO relay system. By assuming a diffuse 2-D field of isotropic scatterers around the receivers, the correlation between the channel coefficients of antennas $1 \le i,j \le M$ of the $k$-th user is modeled as in [28] and is given by
\begin{equation} \label{eq:47}
{\left[ {{{\mathbf{\Theta }}_k}} \right]_{ij}} = \frac{1}{{{\theta _{k,\max }} - {\theta _{k,\min }}}}\int_{{\theta _{k,\min }}}^{{\theta _{k,\max }}} {{e^{i\frac{{2\pi }}{\lambda }{d_{i,j}}\cos \left( \theta  \right)}}d\theta }
\end{equation}
where $\lambda $ denotes the signal wavelength, ${d_{i,j}}$ is the distance between transmit antennas $i$ and $j$, ${\theta _k}$ is the azimuth angle of user $k$ with respect to the orientation perpendicular to the array axis, ${\theta _{k,\max }} - {\theta _{k,\min }}$ indicates the angular spread of departure to user $k$. Other system parameters for performance evaluation are given in Table I.
%\begin{verbatim}
%\tabcap{<table caption width>}
\begin{table}
\renewcommand{\arraystretch}{0.9}
\caption{System Parameters}
\begin{center}
\begin{small}
\begin{tabular}{|| c | c ||}
\hline \hline
%\toprule
Parameters & Value \\
\hline \hline
%\midrule
Channel bandwidth & 10 [MHz] \\
%\hline
Thermal noise density & -174 [dBm] \\
%\hline
\!\!Regularization parameter\!\!  & ${\alpha _1} = {\alpha _2} = {K \mathord{\left/
 {\vphantom {K {\left( {10M\rho } \right)}}} \right.
 \kern-\nulldelimiterspace} {\left( {10M\rho } \right)}}$ \\
%\hline
Path-loss model & $L = 10{\log _{10}}\left( {{{{d^n}} \mathord{\left/
 {\vphantom {{{d^n}} G}} \right.
 \kern-\nulldelimiterspace} G}} \right)$, $d$ in meter \\
 & \!\!$L \!\!=\!\! 128.1 \!+\! 37.6{\log _{10}}D$, $D$ in kilometer\!\! \\
 & $G=0.029512$, $n=3.76$ \\
Distance & $d_{sr}=2500$ [m], $d_{rd}=1500$ [m] \\
\hline \hline
%\bottomrule
\end{tabular}
\end{small}
\end{center}
\end{table}
%\end{verbatim}

Fig. 2 compares the maximum sum rate of M-MIMO-ADF relaying by the MC simulation to that by the deterministic approximation with RZF precoding in correlated channels ($M=768$, $K=64$, ${{\mathbf{\Theta }}_k} \ne {{\mathbf{I}}_M}{\text{ }}\forall k$). From Fig. 2(a) and Fig. 2(b), we observe that for both imperfect CSIT ($\tau _k^2 = 0.1$) and perfect CSIT ($\tau _k^2 = 0$), the deterministic approximation achieves almost the same sum rate as the MC result. From Fig. 2(a), we see that the sum rate of M-MIMO-ADF relaying increases with the growth of transmit power, and it approaches a constant at high SNR region. This is because with the increasing transmit power, the multi-user interference becomes much higher than the noise at the $k$-th user, thus the sum rate of M-MIMO-ADF relaying is limited by the multi-user interferences at high SNR. Besides, we also find that the gap between the sum rate of M-MIMO-DF by MC and that by deterministic approximation is greater than the gap between the sum rate of M-MIMO-AF by MC and that by deterministic approximation, this is due to the fact that the sum rate of M-MIMO-DF relaying by deterministic approximation is an ergodic sum channel capacity in nature, whereas the sum rate of M-MIMO-DF relaying by MC is a minimum sum channel capacity of source-relay and relay-destination channels. From Fig. 2(b), we see that the sum rate of M-MIMO-ADF relaying increases with the growth of SNR, and the sum rate is proportional to the SNR. The reason is that for perfect CSIT ($\tau _k^2 = 0$), the multi-user interference approaches to zero with RZF precoding, and from Eq. (11) and Eq. (17) we find that the sum rate should be proportional to the SNR.
\begin{figure}[!t]
\centering \subfigure[$\tau _k^2 = 0.1$]{\label{Fig2.sub.1}
\includegraphics[width=3.5in]{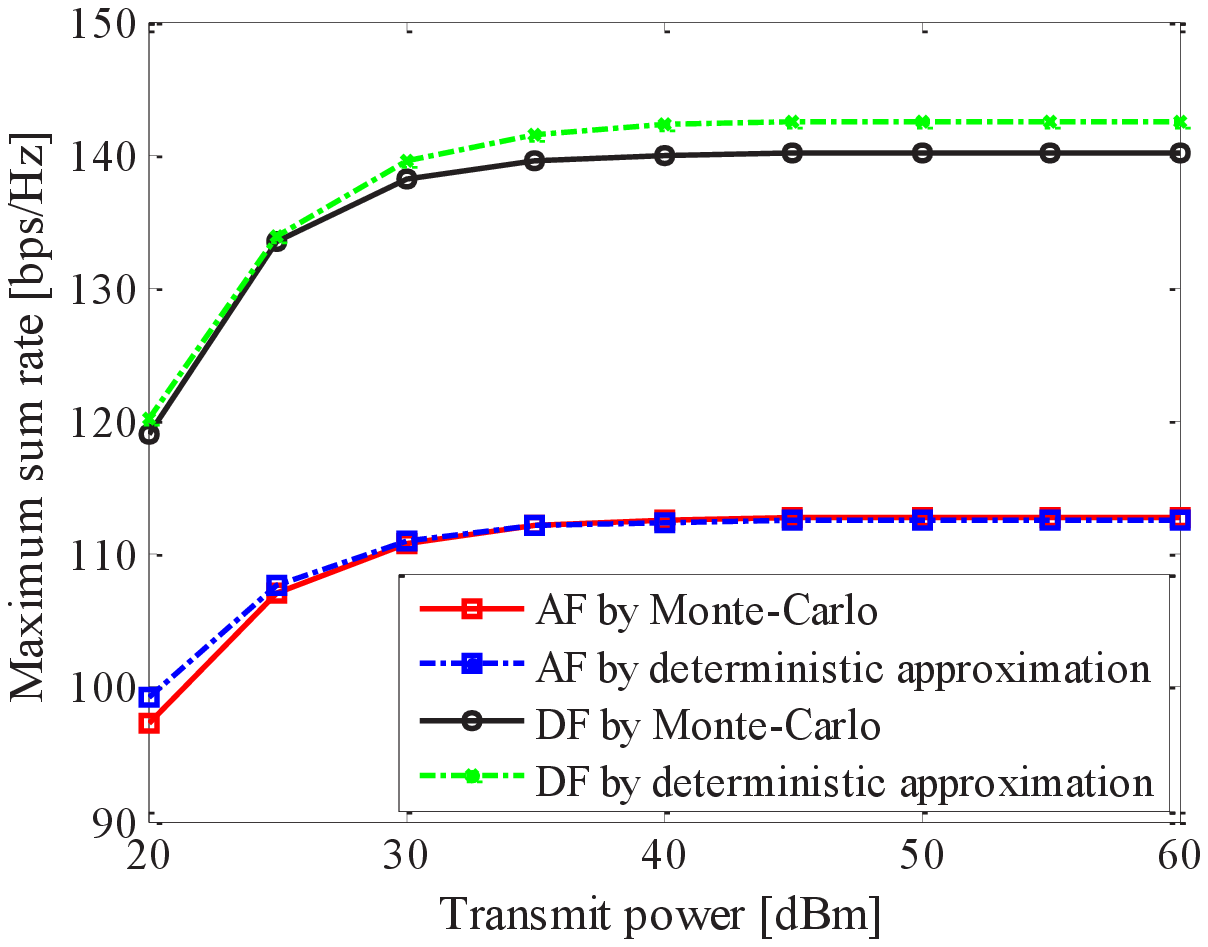}}
\subfigure[$\tau _k^2 = 0$]{ \label{Fig2.sub.2}
\includegraphics[width=3.5in]{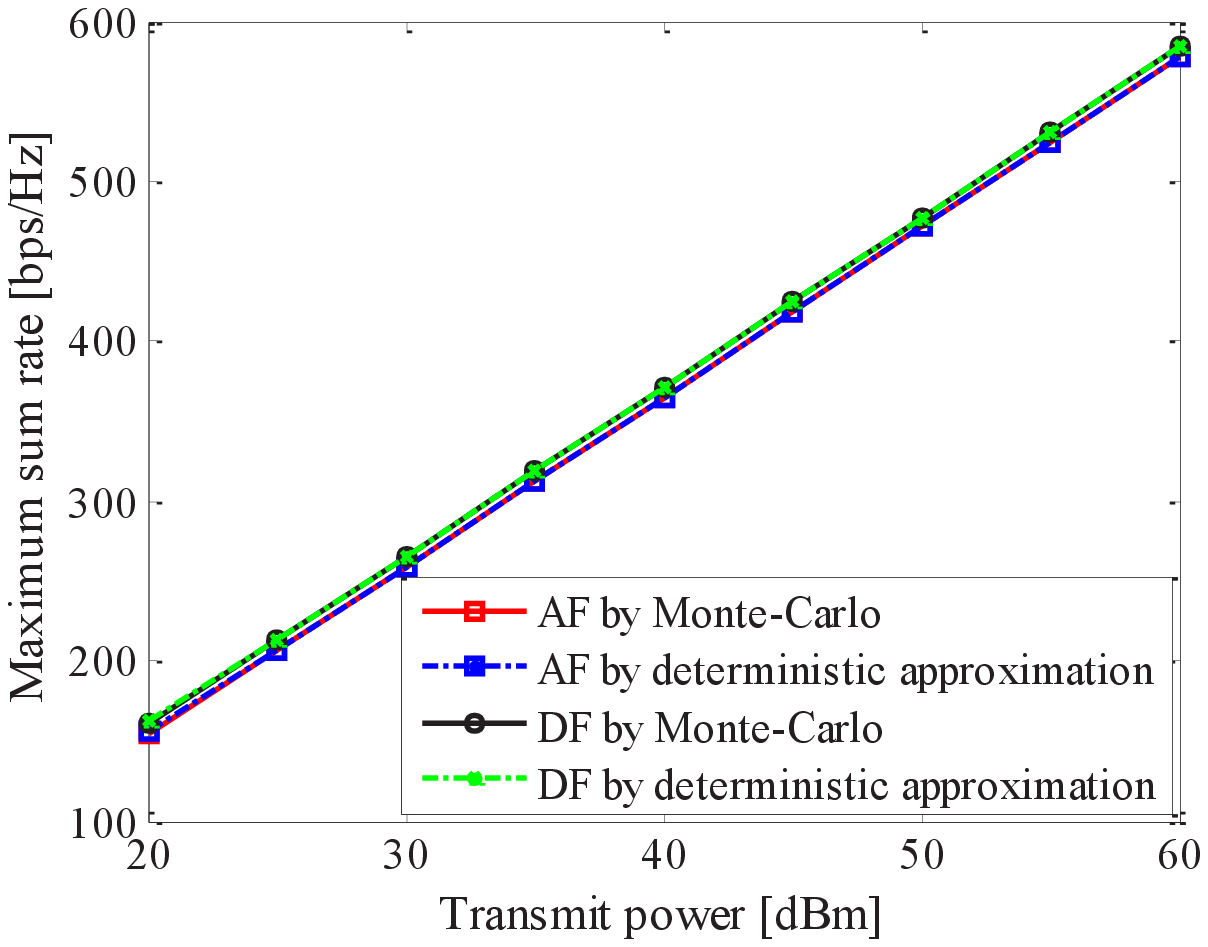}}
\caption{Maximum sum rate versus transmit power with $M=768$, $K=64$, ${{\mathbf{\Theta }}_k} \ne {{\mathbf{I}}_M}\left( {{d_{i,j}} = 0.5\lambda } \right)$.} \label{Fig.2}
\end{figure}

Fig. 3 compares the maximum sum rate of the M-MIMO-ADF relaying by the MC simulation to that by the deterministic approximation with RZF precoding in uncorrelated channels ($M=768$, $K=64$, ${{\mathbf{\Theta }}_k} = {{\mathbf{I}}_M}{\text{ }}\forall k$). It can be observed that for both imperfect CSIT ($\tau _k^2 = 0.1$) and perfect CSIT ($\tau _k^2 = 0$), the deterministic approximation matches well with the MC results. Comparing Fig. 3(a) to Fig. 2(a), it can be found that the maximum sum rate of M-MIMO-ADF relaying for uncorrelated channels is much higher than that for correlated channels.
\begin{figure}[!t]
\centering \subfigure[$\tau _k^2 = 0.1$]{\label{Fig3.sub.1}
\includegraphics[width=3.5in]{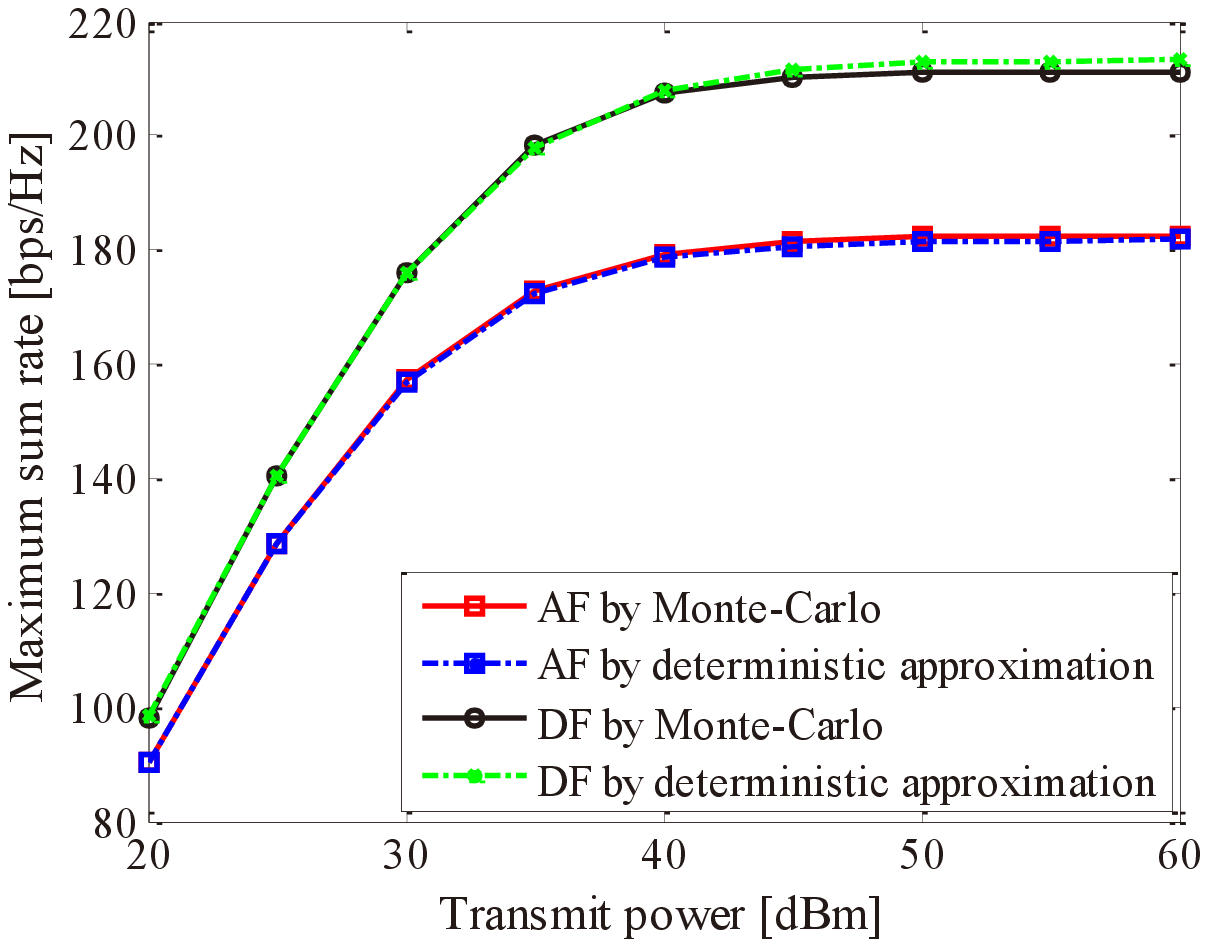}}
\subfigure[$\tau _k^2 = 0$]{ \label{Fig3.sub.2}
\includegraphics[width=3.5in]{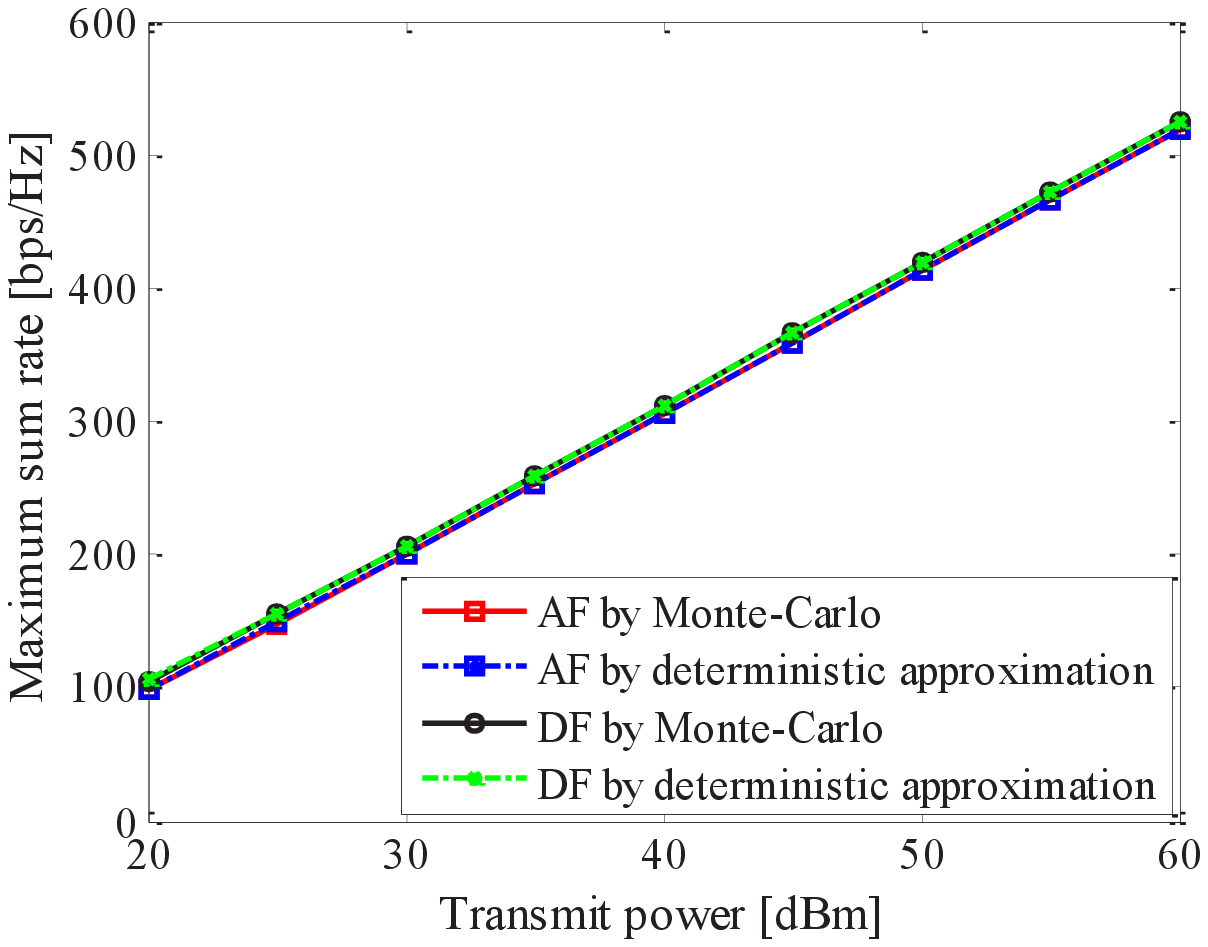}}
\caption{Maximum sum rate versus transmit power with $M=768$, $K=64$, ${{\mathbf{\Theta }}_k} = {{\mathbf{I}}_M}$.} \label{Fig.3}
\end{figure}

Fig. 4 compares the maximum sum rate of the M-MIMO-ADF relaying by the MC simulation to that by the deterministic approximation with RZF precoding in correlated channels ($M=256$, $K=32$, ${{\mathbf{\Theta }}_k} \ne {{\mathbf{I}}_M}{\text{ }}\forall k$). It can be observed that if $M$ and $K$ are reduced to $M=256$ and $K=32$, the deterministic approximation matches well with the MC result for both imperfect CSIT ($\tau _k^2 = 0.1$) and perfect CSIT ($\tau _k^2 = 0$). However, the gap between the MC result and the deterministic approximation becomes larger as compared to that in Fig. 3, since the deterministic approximation by Eq. (19) and Eq. (20) becomes less accurate for a smaller $M$.
\begin{figure}[!t]
\centering \subfigure[$\tau _k^2 = 0.1$]{\label{Fig4.sub.1}
\includegraphics[width=3.5in]{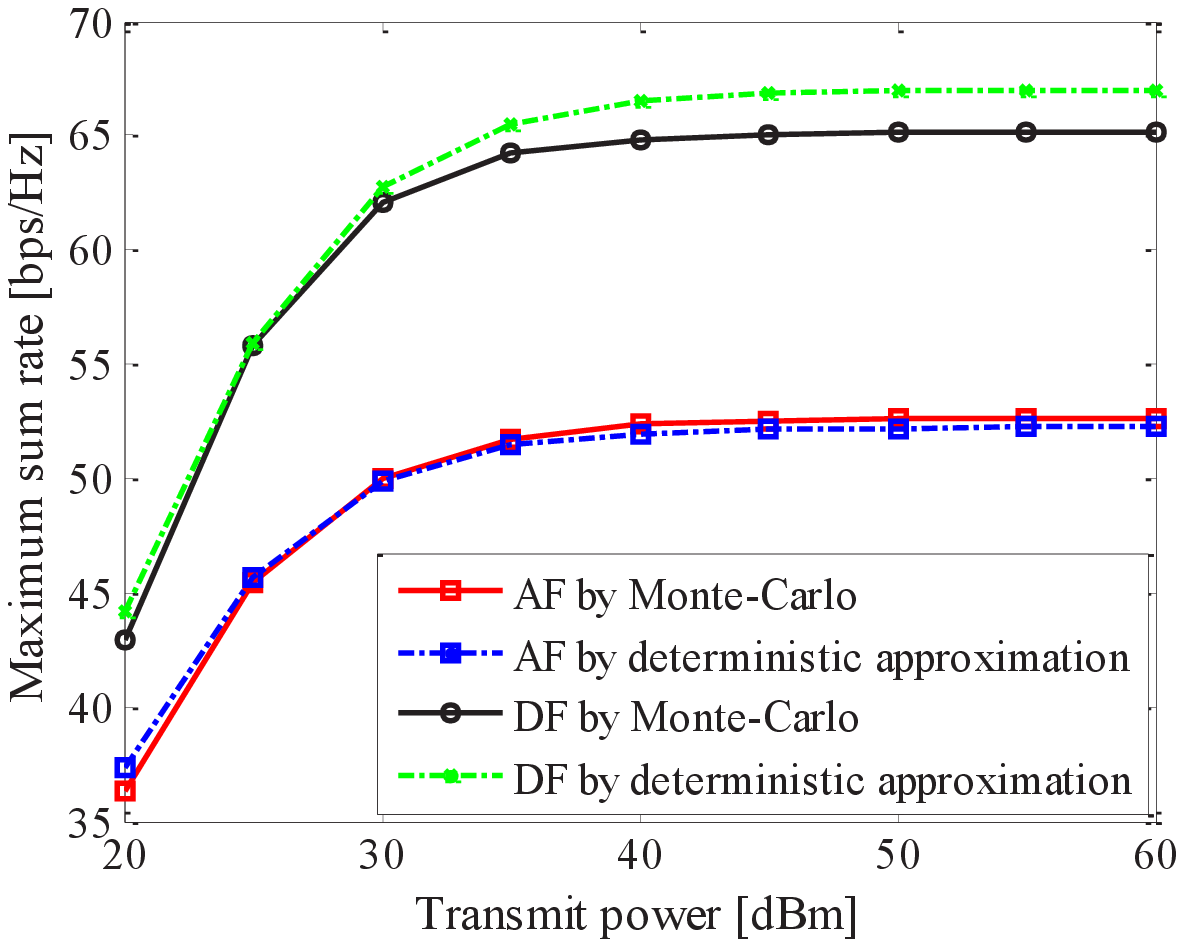}}
\subfigure[$\tau _k^2 = 0$]{ \label{Fig4.sub.2}
\includegraphics[width=3.5in]{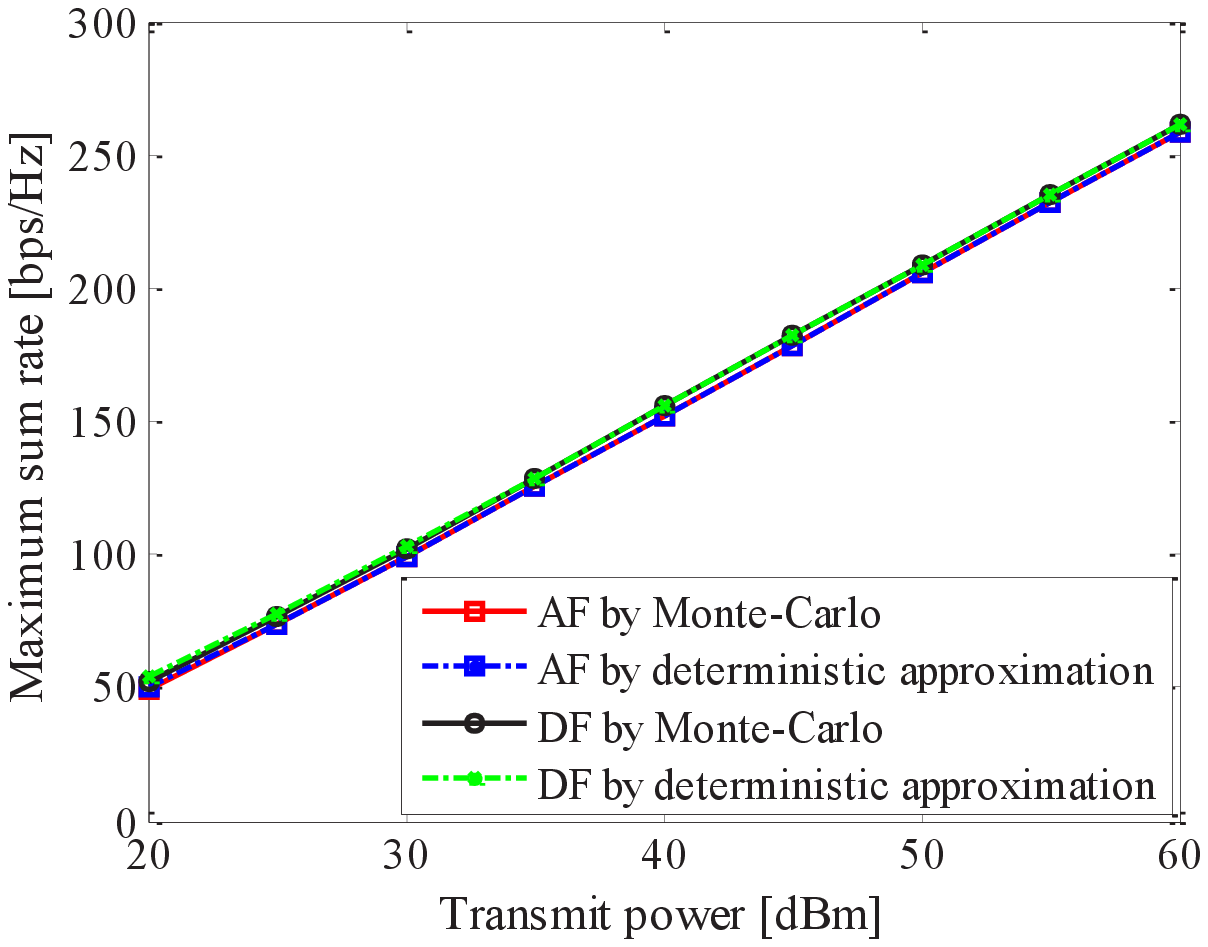}}
\caption{Maximum sum rate versus transmit power with $M=256$, $K=32$, ${{\mathbf{\Theta }}_k} \ne {{\mathbf{I}}_M}\left( {{d_{i,j}} = 0.5\lambda } \right)$.} \label{Fig.4}
\end{figure}

\begin{figure}[!t]
\centering \subfigure[$\tau _k^2 = 0.1$]{\label{Fig5.sub.1}
\includegraphics[width=3.5in]{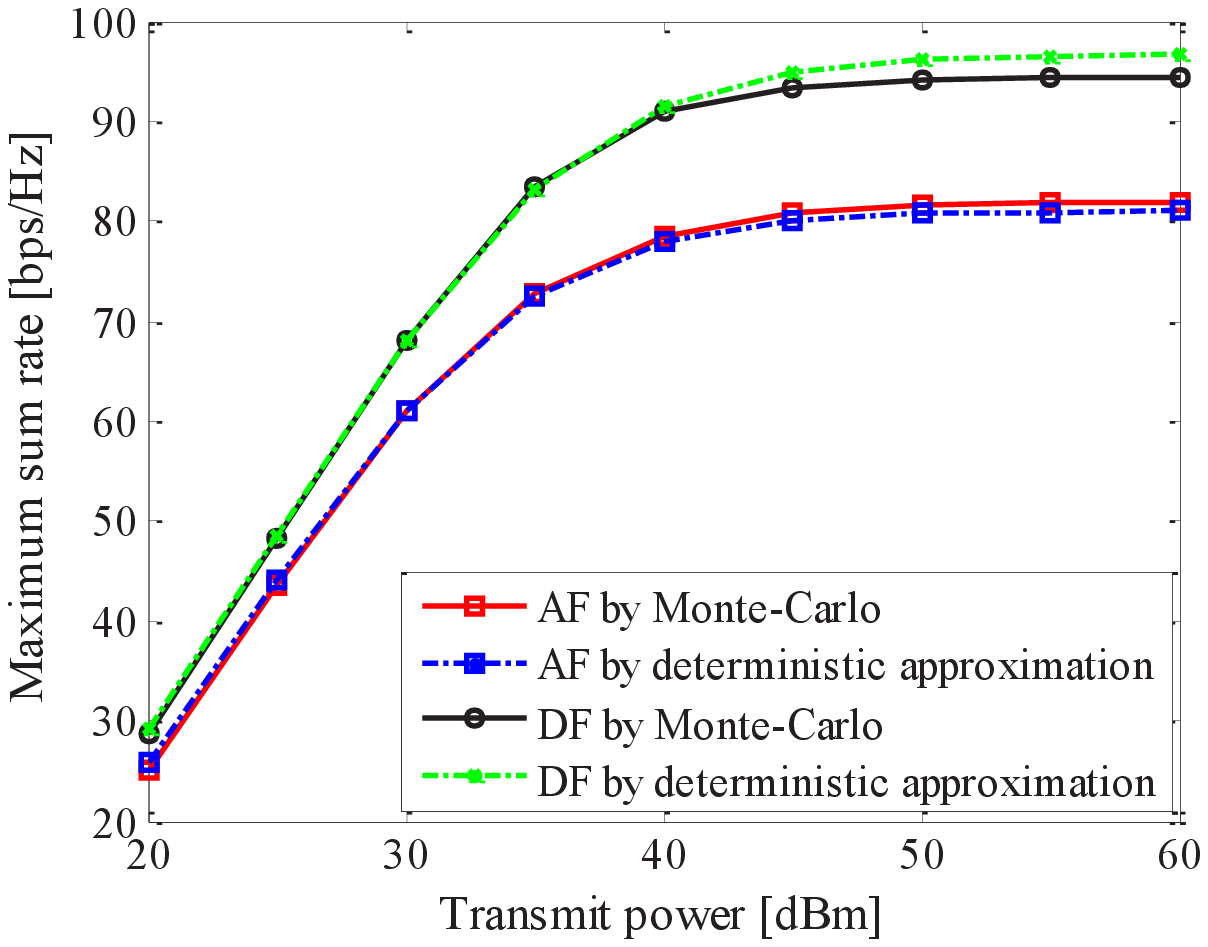}}
\subfigure[$\tau _k^2 = 0$]{ \label{Fig5.sub.2}
\includegraphics[width=3.5in]{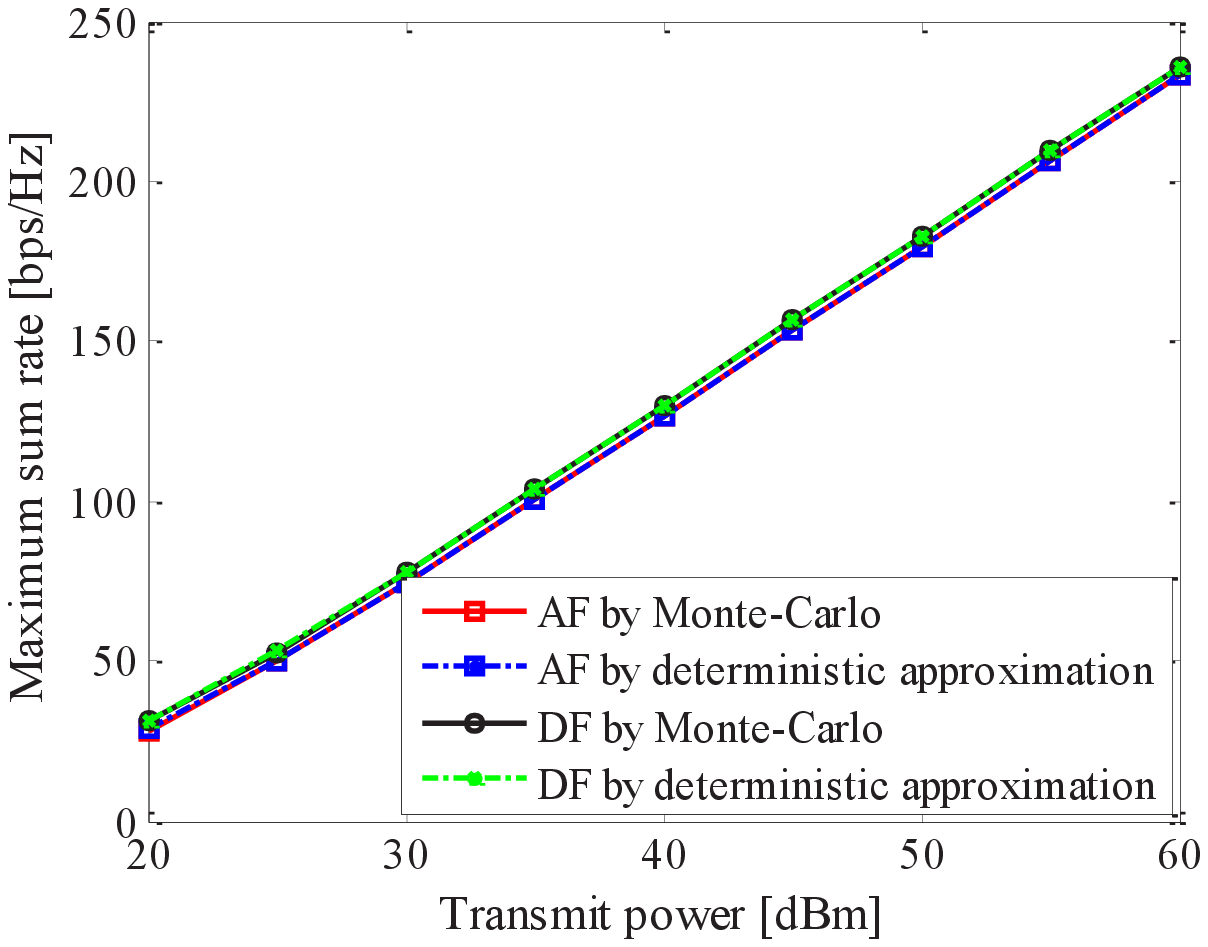}}
\caption{Maximum sum rate versus transmit power with $M=256$, $K=32$, ${{\mathbf{\Theta }}_k} = {{\mathbf{I}}_M}$.} \label{Fig.5}
\end{figure}

Fig. 5 compares the maximum sum rate of the M-MIMO-ADF relaying by the MC simulation to that by deterministic approximation with RZF precoding in uncorrelated channels ($M=256$, $K=32$, ${{\mathbf{\Theta }}_k} = {{\mathbf{I}}_M}{\text{ }}\forall k$). It can be observed that if $M$ and $K$ are reduced to $M=256$ and $K=32$, the deterministic approximation matches well with the MC result for both imperfect CSIT ($\tau _k^2 = 0.1$) and perfect CSIT ($\tau _k^2 = 0$). Comparing Fig. 5(a) to Fig. 4(a), we find that the maximum sum rate of M-MIMO-ADF relaying with RZF precoding in uncorrelated channels is much higher than that in correlated channels.

From Fig. 2 - Fig. 5 we conclude that the deterministic approximation for the received SINR at user $k$ in the M-MIMO-ADF relaying and the results of Theorem 1, Theorem 2, Proposition 1 and Corollary 1 are accurate.

\begin{figure}[!t]
\centering \subfigure[$\tau _k^2 = 0.1$]{\label{Fig6.sub.1}
\includegraphics[width=3.5in]{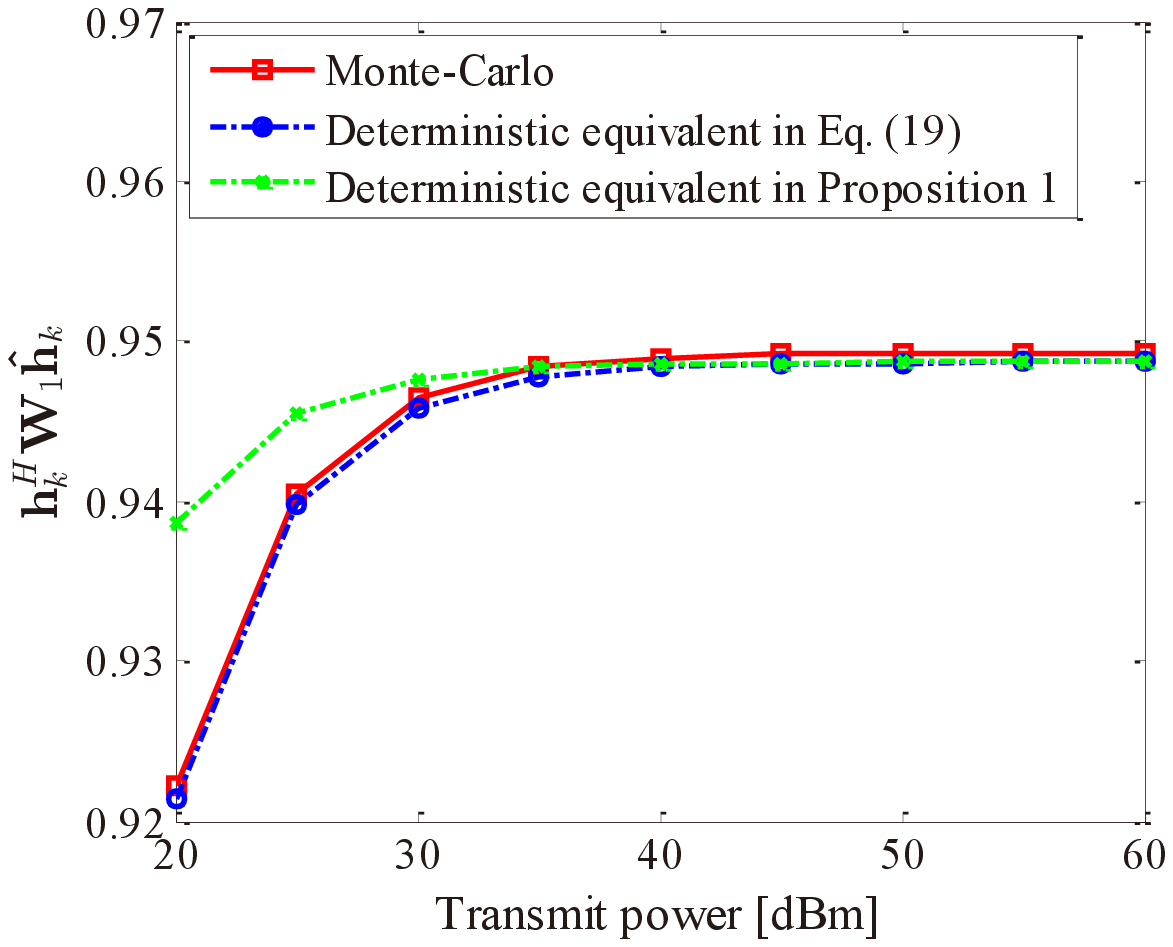}}
\subfigure[$\tau _k^2 = 0$]{ \label{Fig6.sub.2}
\includegraphics[width=3.5in]{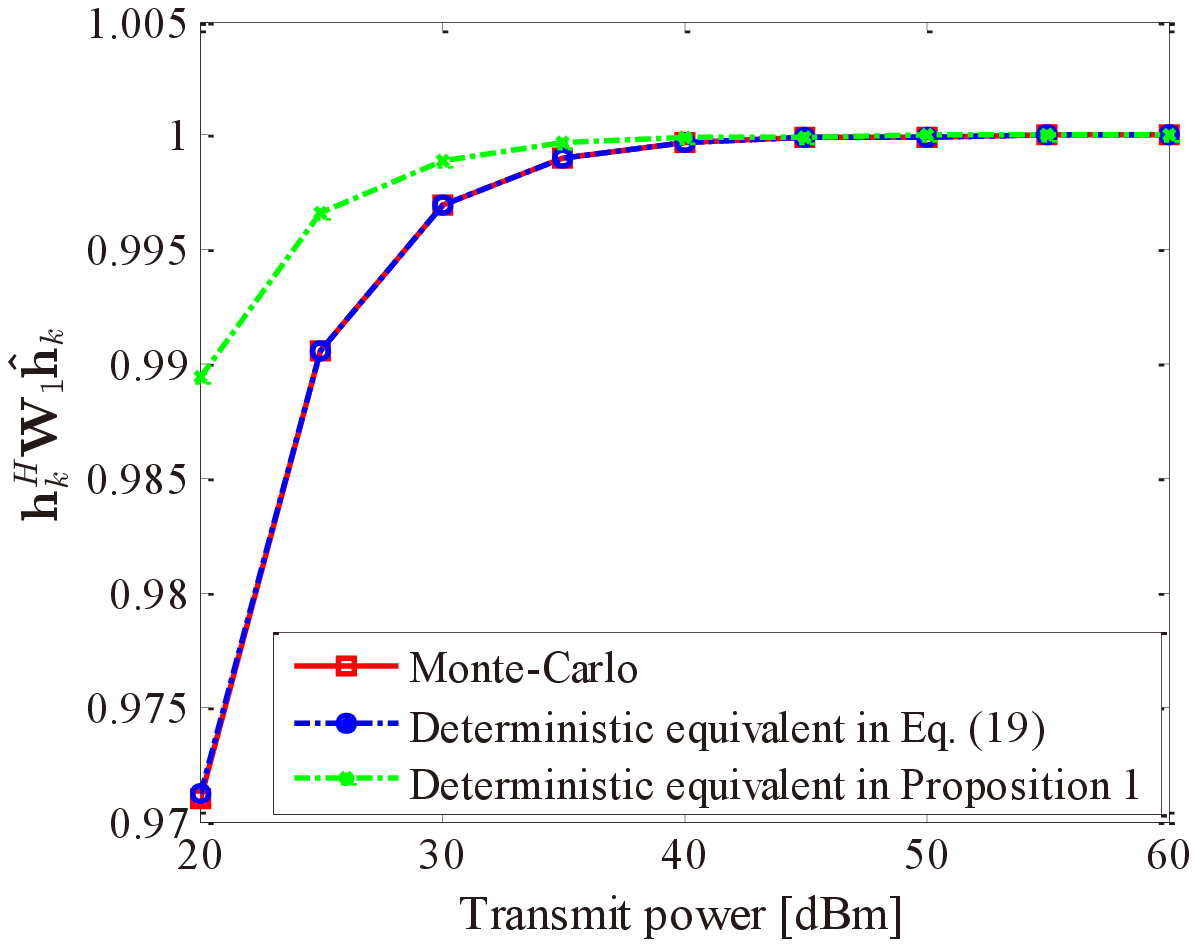}}
\caption{${\mathbf{h}}_k^H{{\mathbf{W}}_1}{{\mathbf{\hat h}}_k}$ versus transmit power with $M=768$, $K=64$, ${{\mathbf{\Theta }}_k} \ne {{\mathbf{I}}_M}\left( {{d_{i,j}} = 0.5\lambda } \right)$.} \label{Fig.6}
\end{figure}

\begin{figure}[!t]
\centering \subfigure[$\tau _k^2 = 0.1$]{\label{Fig7.sub.1}
\includegraphics[width=3.5in]{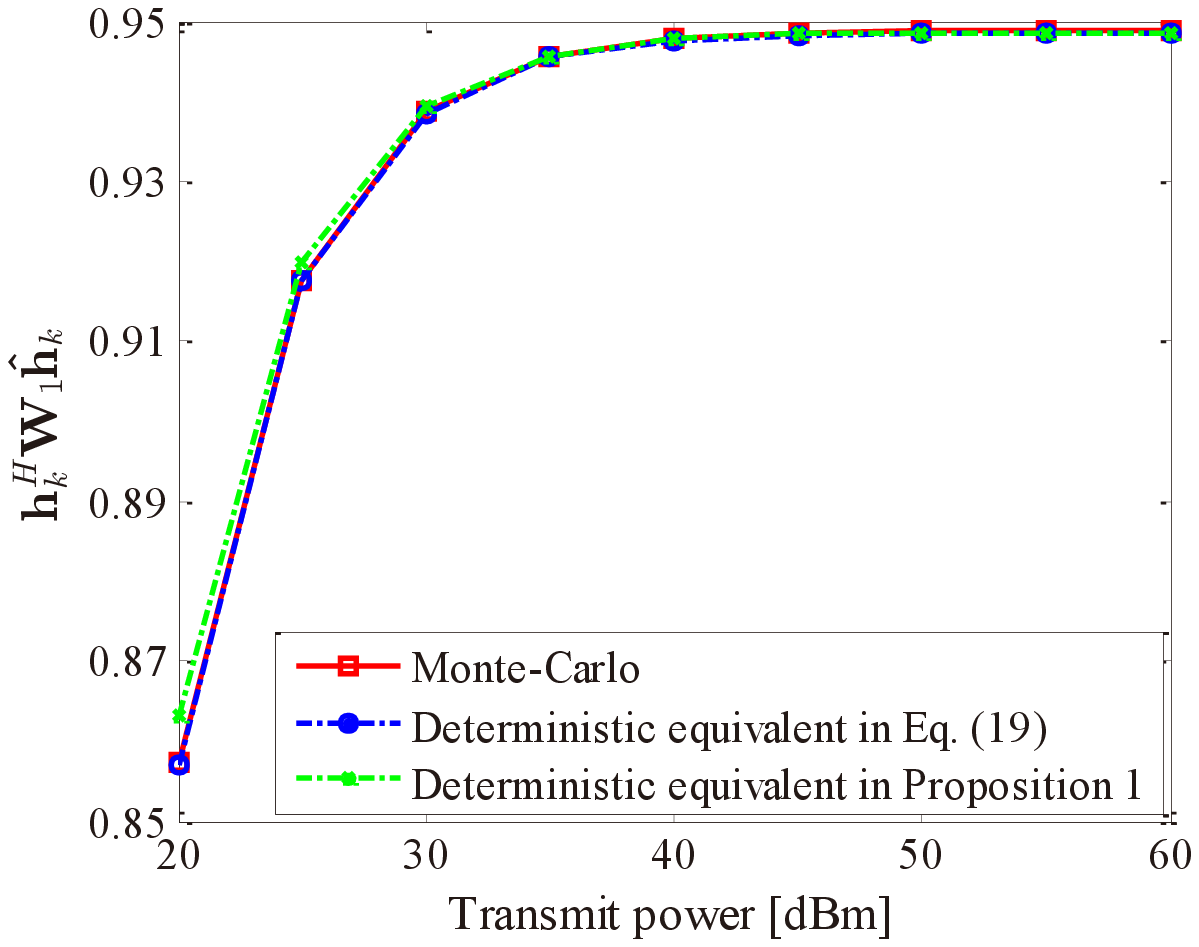}}
\subfigure[$\tau _k^2 = 0$]{ \label{Fig7.sub.2}
\includegraphics[width=3.5in]{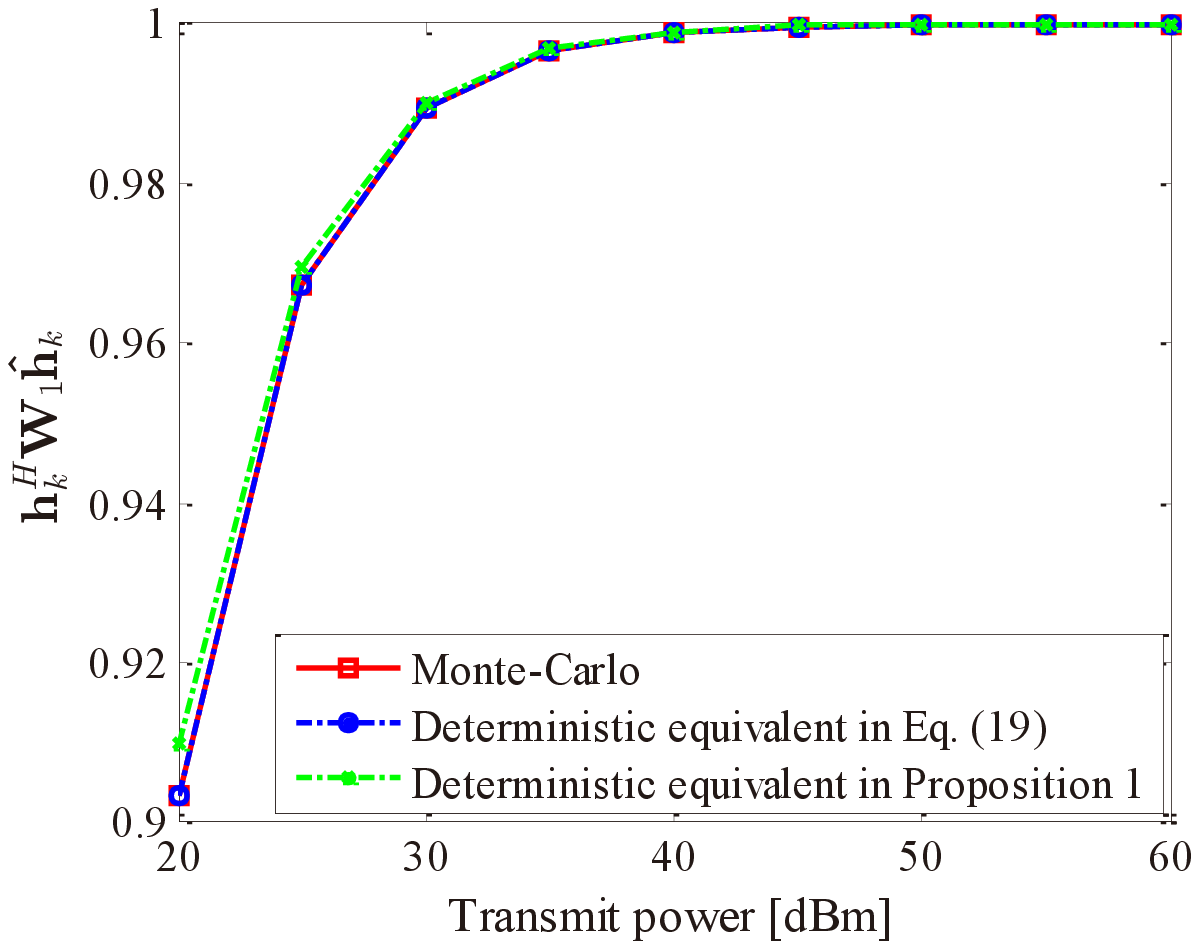}}
\caption{${\mathbf{h}}_k^H{{\mathbf{W}}_1}{{\mathbf{\hat h}}_k}$ versus transmit power with $M=768$, $K=64$, ${{\mathbf{\Theta }}_k} = {{\mathbf{I}}_M}$.} \label{Fig.7}
\end{figure}
Fig. 6 and Fig. 7 compare the MC result of ${\mathbf{h}}_k^H{{\mathbf{W}}_1}{{\mathbf{\hat h}}_k}$ to the deterministic equivalent in Eq. (19) of [21] and the deterministic equivalent in Proposition 1 for the M-MIMO-ADF relaying with RZF precoding in both correlated channels ($M=768$, $K=64$, ${{\mathbf{\Theta }}_k} \ne {{\mathbf{I}}_M}{\text{ }}\forall k$) and uncorrelated channels ($M=768$, $K=64$, ${{\mathbf{\Theta }}_k} = {{\mathbf{I}}_M}{\text{ }}\forall k$). We find that for both imperfect CSIT ($\tau _k^2 = 0.1$) and perfect CSIT ($\tau _k^2 = 0$), the deterministic equivalent in Eq. (19) and that in Proposition 1 both match well with the MC result, and at low SNR region the matching for the deterministic equivalent in Proposition 1 is a little worse than that in Eq. (19).

From Fig. 6 and Fig. 7, we conclude that the deterministic equivalent in Proposition 1 matches well with the MC result of ${\mathbf{h}}_k^H{{\mathbf{W}}_1}{{\mathbf{\hat h}}_k}$ for both correlated and uncorrelated channels if $M,K \to \infty $ and $M \gg K$.

\section{Conclusion}
In this paper we studied a massive MIMO relay system with linear precoding and analyzed the system performance for a large number of antennas and users. Our channel model is realistic as the channels are assumed to have imperfect CSIT and per-user channel correlation, and our source-relay channels are massive MIMO ones rather than massive MISO ones. We use large random matrix theory (RMT) to derive the deterministic equivalent of the SINR at each user, as the number of transmit antennas and the users $M,K \to \infty $ and $M \gg K$. Simulation results have shown that the deterministic equivalent of the SINR at each user in M-MIMO-ADF relaying and the results of Theorem 1, Theorem 2, Proposition 1 and Corollary 1 are accurate.

As in practice individual channel gains may not be available for massive MIMO systems due to some reasons such as limited feedback, pilot contamination, time-sensitive applications and so on, our large system performance approximations can be applied to simulate the system behavior without having to carry out extensive MC simulations, and they can be used to solve practical optimization problems. However, a challenging problem still remains for future research on how to accurately calculate the deterministic equivalent of the interference to each user if $M,K \to \infty $ and the value of $K$ is close to $M$.

\appendices
\section{Proof of Theorem 1}
Without loss of generality, let's first prove $\left| {{{\mathbf{h}}_k^H}{{{\mathbf{\hat W}}}_1}{{{\mathbf{\hat h}}}_k}} \right| \gg \left| {{{\mathbf{h}}_k^H}{{{\mathbf{\hat W}}}_1}{{{\mathbf{\hat h}}}_{k'}}} \right|$. As we know, ${{\mathbf{\hat W}}_1} = {\left( {{\mathbf{\hat H}}_1^H{{{\mathbf{\hat H}}}_1} + M{\alpha _1}{{\mathbf{I}}_M}} \right)^{ - 1}}$ is a symmetric and positive definite matrix, it can be decomposed as ${{\mathbf{\hat W}}_1} = {\mathbf{U}}_1^H{{\mathbf{\Lambda }}_1}{{\mathbf{U}}_1}$, where ${{\mathbf{\Lambda }}_1} = diag\left( {{\lambda _1}, \cdots ,{\lambda _M}} \right)$ is a $M \times M$ diagonal matrix containing $M$ positive eigenvalues of ${{\mathbf{\hat W}}_1}$. As
\begin{eqnarray} \label{eq:48}
  {\mathbf{\hat W}}_1^{ - 1} &\!\!=\!\!& \left[ {{{{\mathbf{\hat h}}}_{sr,1}}, \cdots ,{{{\mathbf{\hat h}}}_{sr,K}}} \right]{\left[ {{{{\mathbf{\hat h}}}_{sr,1}}, \cdots ,{{{\mathbf{\hat h}}}_{sr,K}}} \right]^H} + M{\alpha _1}{{\mathbf{I}}_M} \hfill \nonumber \\
   &\!\!=\!\!& {\mathbf{U}}_1^H\left( {{{{\mathbf{\Lambda '}}}_1} + M{\alpha _1}{{\mathbf{I}}_M}} \right){{\mathbf{U}}_1}
\end{eqnarray}
where ${{\mathbf{\Lambda '}}_1} = diag\left( {\underbrace {{{\lambda '}_1}, \cdots ,{{\lambda '}_K}}_K,\underbrace {0, \cdots ,0}_{M - K}} \right)$, ${\lambda '_m} > 0$ for $\forall m \in \left\{ {1, \cdots ,K} \right\}$. Then ${{\mathbf{\hat W}}_1} = {\left( {{\mathbf{\hat H}}_1^H{{{\mathbf{\hat H}}}_1} + M{\alpha _1}{{\mathbf{I}}_M}} \right)^{ - 1}}$ can be decomposed as ${{\mathbf{\hat W}}_1} = {\mathbf{U}}_1^H{{\mathbf{\Lambda }}_1}{{\mathbf{U}}_1}$, where ${{\mathbf{\Lambda }}_1} = diag\left( {\lambda _1}, \cdots ,{\lambda _K}, {\lambda _{K + 1}} \cdots ,{\lambda _M} \right)
$ is a $M \times M$ diagonal matrix containing $M$ positive eigenvalues of ${{\mathbf{\hat W}}_1}$, ${\lambda _m} = {\left( {{{\lambda '}_m} + M{\alpha _1}} \right)^{ - 1}} > 0$ for $m \in \left\{ {1, \cdots ,K} \right\}$ and ${\lambda _m} = {\left( {M{\alpha _1}} \right)^{ - 1}} > 0$ for $m \in \left\{ {K + 1, \cdots ,M} \right\}$.

Let ${{\mathbf{U}}_1}{\mathbf{\Theta }}_k^{{1 \mathord{\left/
 {\vphantom {1 2}} \right.
 \kern-\nulldelimiterspace} 2}}{{\mathbf{z}}_k} \triangleq {\left[ {{x_1},{x_2}, \cdots ,{x_M}} \right]^T}$, ${{\mathbf{U}}_1}{\mathbf{\Theta }}_{k'}^{{1 \mathord{\left/
 {\vphantom {1 2}} \right.
 \kern-\nulldelimiterspace} 2}}{{\mathbf{z}}_{k'}} \triangleq {\left[ {{y_1},{y_2}, \cdots ,{y_M}} \right]^T}$ where ${x_m}$ and ${y_m}$ ($m \in \left\{ {1,2, \cdots M} \right\}$) follows the same distribution, if the accuracy parameter of the channel estimate satisfies the following condition as $\tau  = {\tau _k} = {\tau _{k'}}$, then we have
\begin{eqnarray} \label{eq:49}
  &&\!\!\!\!\!\!\!\!\!\!\!\! {\mathbf{h}}_k^H{{{\mathbf{\hat W}}}_1}{{{\mathbf{\hat h}}}_k}  \xrightarrow{{M \to \infty }}  {\mathbf{h}}_k^H{\mathbf{U}}_1^H{{\mathbf{\Lambda }}_1}{{\mathbf{U}}_1}{{{\mathbf{\hat h}}}_k} \hfill \nonumber \\
  && \qquad\; \xrightarrow{{M \to \infty }}  \! M\!\left( {1 \!-\! {\tau ^2}} \right)\!{\mathbf{z}}_k^H{\left(\! {{\mathbf{\Theta }}_k^{{1 \mathord{\left/
 {\vphantom {1 2}} \right.
 \kern-\nulldelimiterspace} 2}}} \!\right)^{\!H}}\!\!{\mathbf{U}}_1^H{{\mathbf{\Lambda }}_1}{{\mathbf{U}}_1}{\mathbf{\Theta }}_k^{{1 \mathord{\left/
 {\vphantom {1 2}} \right.
 \kern-\nulldelimiterspace} 2}}{{\mathbf{z}}_k} \nonumber \\
 && \qquad\qquad\; + M\tau \sqrt {1 \!-\! {\tau ^2}} {\mathbf{z}}_k^H{\left( \!{{\mathbf{\Theta }}_k^{{1 \mathord{\left/
 {\vphantom {1 2}} \right.
 \kern-\nulldelimiterspace} 2}}} \!\right)^{\!H}}{\!\!\mathbf{U}}_1^H{{\mathbf{\Lambda }}_1}{{\mathbf{U}}_1}{\mathbf{\Theta }}_k^{{1 \mathord{\left/
 {\vphantom {1 2}} \right.
 \kern-\nulldelimiterspace} 2}}{{\mathbf{q}}_k} \hfill \nonumber \\
 && \qquad\; \xrightarrow{{M \to \infty }}  \! M \! \left( {1 \!-\! {\tau ^2}}  \right)\!\sum\limits_{m = 1}^M {{\lambda _m}{{\left| {{x_m}} \right|}^2}}  \nonumber \\
 && \qquad\qquad\; + M\tau \sqrt {1 \!-\! {\tau ^2}} {\mathbf{z}}_k^H{\left(\! {{\mathbf{\Theta }}_k^{{1 \mathord{\left/
 {\vphantom {1 2}} \right.
 \kern-\nulldelimiterspace} 2}}} \!\right)^H}{\mathbf{U}}_1^H{{\mathbf{\Lambda }}_1}{{\mathbf{U}}_1}{\mathbf{\Theta }}_k^{{1 \mathord{\left/
 {\vphantom {1 2}} \right.
 \kern-\nulldelimiterspace} 2}}{{\mathbf{q}}_k} \nonumber \\
\end{eqnarray}
\begin{eqnarray} \label{eq:50}
  &&\!\!\!\!\!\!\!\!\!\!\!\! {\mathbf{h}}_k^H{{{\mathbf{\hat W}}}_1}{{{\mathbf{\hat h}}}_{k'}}  \xrightarrow{{M \to \infty }}  {\mathbf{h}}_k^H{\mathbf{U}}_1^H{{\mathbf{\Lambda }}_1}{{\mathbf{U}}_1}{{{\mathbf{\hat h}}}_{k'}} \hfill \nonumber \\
  && \qquad\;\;\,  \xrightarrow{{M \to \infty }}  \! M  \! \left( {1 \!-\! {\tau ^2}} \right){\!\mathbf{z}}_k^H{\! \left(\! {{\mathbf{\Theta }}_k^{{1 \mathord{\left/
 {\vphantom {1 2}} \right.
 \kern-\nulldelimiterspace} 2}}} \!\right)^H}\!\!{\mathbf{U}}_1^H{{\mathbf{\Lambda }}_1}{{\!\mathbf{U}}_1}{\!\mathbf{\Theta }}_{k'}^{{1 \mathord{\left/
 {\vphantom {1 2}} \right.
 \kern-\nulldelimiterspace} 2}}{{\mathbf{z}}_{k'}} \nonumber \\
 && \qquad\qquad\; + M\tau \sqrt {1 \!-\! {\tau ^2}} {\mathbf{z}}_k^H{\! \left(\! {{\mathbf{\Theta }}_k^{{1 \mathord{\left/
 {\vphantom {1 2}} \right.
 \kern-\nulldelimiterspace} 2}}} \!\right)^H}\!\!{\mathbf{U}}_1^H{{\mathbf{\Lambda }}_1}{{\!\mathbf{U}}_1}{\!\mathbf{\Theta }}_{k'}^{{1 \mathord{\left/
 {\vphantom {1 2}} \right.
 \kern-\nulldelimiterspace} 2}}{{\mathbf{q}}_{k'}} \hfill \nonumber \\
 && \qquad\;\;\,  \xrightarrow{{M \to \infty }}  \! M \! \left( {1 - {\tau ^2}} \right)\sum\limits_{m = 1}^M {{\lambda _m}x_m^H{y_m}} \nonumber \\
 && \qquad\qquad\; + M\tau \sqrt {1 \!-\! {\tau ^2}} {\mathbf{z}}_k^H{\left( {{\mathbf{\Theta }}_k^{{1 \mathord{\left/
 {\vphantom {1 2}} \right.
 \kern-\nulldelimiterspace} 2}}} \right)^H}{\mathbf{U}}_1^H{{\mathbf{\Lambda }}_1}{{\mathbf{U}}_1}{\mathbf{\Theta }}_{k'}^{{1 \mathord{\left/
 {\vphantom {1 2}} \right.
 \kern-\nulldelimiterspace} 2}}{{\mathbf{q}}_{k'}} \nonumber \\
\end{eqnarray}
As the second terms in the above formulas satisfy the following condition
\begin{eqnarray} \label{eq:51}
&&\!\!\!\!\!\!\!\!\!\!\!\! {\mathbf{z}}_k^H{\!\left(\! {{\mathbf{\Theta }}_k^{{1 \mathord{\left/
 {\vphantom {1 2}} \right.
 \kern-\nulldelimiterspace} 2}}} \!\right)^{\!H}}\!{\mathbf{U}}_1^H{{\mathbf{\Lambda }}_1}\!{{\mathbf{U}}_1}{\mathbf{\Theta }}_k^{{1 \mathord{\left/
 {\vphantom {1 2}} \right.
 \kern-\nulldelimiterspace} 2}}\!{{\mathbf{q}}_k} - {\mathbf{z}}_k^H{\!\left(\! {{\mathbf{\Theta }}_k^{{1 \mathord{\left/
 {\vphantom {1 2}} \right.
 \kern-\nulldelimiterspace} 2}}} \!\right)^{\!H}}\!{\mathbf{U}}_1^H{{\mathbf{\Lambda }}_1}\!{{\mathbf{U}}_1}{\mathbf{\Theta }}_{k'}^{{1 \mathord{\left/
 {\vphantom {1 2}} \right.
 \kern-\nulldelimiterspace} 2}}\!{{\mathbf{q}}_{k'}} \nonumber \\
 &&\!\!\!\!\!\!\!\!\!\!\!\! \xrightarrow{{M \to \infty }}0
\end{eqnarray}
and ${\lambda _m} > 0$, $E\left( {{x_m}} \right) = E\left( {{y_m}} \right) = 0$, which means that ${\lambda _m}x_m^H{y_m}$ has an equal probability to be positive or negative, while ${\lambda _m}{\left| {{x_m}} \right|^2}$ is always positive. From a statistical point of view, the values of ${\lambda _m}\left| {x_m^H{y_m}} \right|$ and ${\lambda _m}{\left| {{x_m}} \right|^2}$ are the same since ${x_m}$ and ${y_m}$ follows the same distribution. Therefore the first terms in Eq. (49) and Eq. (50) satisfy the following inequality
\begin{equation} \label{eq:52}
\sum\limits_{m = 1}^M {{\lambda _m}{{\left| {{x_m}} \right|}^2}}  - \sum\limits_{m = 1}^M {{\lambda _m}\left| {x_m^H{y_m}} \right|}  \gg 0,
\end{equation}
for $M \to \infty $, then we have $\left| {{\mathbf{h}}_k^H{{{\mathbf{\hat W}}}_1}{{{\mathbf{\hat h}}}_k}} \right| \gg \left| {{\mathbf{h}}_k^H{{{\mathbf{\hat W}}}_1}{\mathbf{\hat h}}_{k'}^H} \right|$.

Similarly, we can also prove $\left| {{\mathbf{h}}_k^H{{{\mathbf{\hat W}}}_2}{{{\mathbf{\hat h}}}_k}} \right| \gg \left| {{\mathbf{h}}_k^H{{{\mathbf{\hat W}}}_2}{{{\mathbf{\hat h}}}_{k'}}} \right|$. So far we have proved that $\left| {{\mathbf{h}}_k^H{{{\mathbf{\hat W}}}_l}{{{\mathbf{\hat h}}}_k}} \right| \gg \left| {{\mathbf{h}}_k^H{{{\mathbf{\hat W}}}_l}{{{\mathbf{\hat h}}}_{k'}}} \right|$ for $l \in \left\{ {1,2} \right\}$, the proof for Theorem 1 is complete.

\section{Proof of Theorem 2}
Without loss of generality, let's first prove $\left| {{\mathbf{\hat h}}_k^H{\mathbf{\hat W}}_1^2{{{\mathbf{\hat h}}}_k}} \right| \!\gg\! \left| {{\mathbf{\hat h}}_k^H{\mathbf{\hat W}}_1^2{{{\mathbf{\hat h}}}_{k'}}} \right|$. If ${{\mathbf{\hat W}}_1} \!=\! {\left( {{\mathbf{\hat H}}_1^H{{{\mathbf{\hat H}}}_1} + M{\alpha _1}{{\mathbf{I}}_M}} \right)^{\!{ - 1}}}$ can be decomposed as ${{\mathbf{\hat W}}_1} = {\mathbf{U}}_1^H{{\mathbf{\Lambda }}_1}{{\mathbf{U}}_1}$ where ${{\mathbf{\Lambda }}_1} = diag\left( {{\lambda _1}, \cdots ,{\lambda _M}} \right)$ is a $M \times M$ diagonal matrix containing $M$ positive eigenvalues of ${{\mathbf{\hat W}}_1}$, ${\lambda _m} > 0$ for $m \in \left\{ {1, \cdots ,K} \right\}$ and ${\lambda _m} = {\left( {M{\alpha _1}} \right)^{ - 1}} > 0$ for $m \in \left\{ {K + 1, \cdots ,M} \right\}$, then ${\mathbf{\hat W}}_1^2 = {\left( {{\mathbf{\hat H}}_1^H{{{\mathbf{\hat H}}}_1} + M{\alpha _1}{{\mathbf{I}}_M}} \right)^{ - 2}}$ is also a symmetric and positive definite matrix and can be decomposed as ${\mathbf{\hat W}}_1^2 = {\mathbf{U}}_1^H{\mathbf{\Lambda }}_1^2{{\mathbf{U}}_1}$, where ${\mathbf{\Lambda }}_1^2 = diag\left( {\lambda _1^2, \cdots ,\lambda _M^2} \right)$.

Let ${{\mathbf{U}}_1}{{\mathbf{\hat h}}_k} \triangleq {\left[ {{x_1},{x_2}, \cdots ,{x_M}} \right]^T}$, ${{\mathbf{U}}_1}{{\mathbf{\hat h}}_{k'}} \triangleq  {\left[ {{y_1},{y_2}, \cdots ,{y_M}} \right]^T}$ where ${x_m}$ and ${y_m}$ ($m \! \in \! \left\{ {1,2, \cdots M} \right\}$) follows the same distribution, then we have
\begin{eqnarray} \label{eq:53}
  {\mathbf{\hat h}}_k^H{\mathbf{\hat W}}_1^2{{{\mathbf{\hat h}}}_k} & \xrightarrow{{M \to \infty }} & {\mathbf{\hat h}}_k^H{\mathbf{U}}_1^H{\mathbf{\Lambda }}_1^2{{\mathbf{U}}_1}{{{\mathbf{\hat h}}}_k} \hfill \nonumber \\
 & \xrightarrow{{M \to \infty }} & \sum\limits_{m = 1}^M {\lambda _m^2{{\left| {{x_m}} \right|}^2}}  \hfill
\end{eqnarray}
\begin{eqnarray} \label{eq:54}
  {\mathbf{\hat h}}_k^H{\mathbf{\hat W}}_1^2{{{\mathbf{\hat h}}}_{k'}} & \xrightarrow{{M \to \infty }} & {\mathbf{\hat h}}_k^H{\mathbf{U}}_1^H{\mathbf{\Lambda }}_1^2{{\mathbf{U}}_1}{{{\mathbf{\hat h}}}_{k'}} \hfill \nonumber \\
  & \xrightarrow{{M \to \infty }} & \sum\limits_{m = 1}^M {\lambda _m^2x_m^H{y_m}}  \hfill
\end{eqnarray}
As $\lambda _m^2 > 0$ and $E\left( {{x_m}} \right) = E\left( {{y_m}} \right) = 0$, which means that both the real and the imaginary part of $\lambda _m^2x_m^H{y_m}$ have an equal probability to be positive or negative, while $\lambda _m^2{\left| {{x_m}} \right|^2}$ is always positive. From a statistical point of view, the values of $\lambda _m^2\left| {x_m^H{y_m}} \right|$ and $\lambda _m^2{\left| {{x_m}} \right|^2}$ are the same since ${x_m}$ and ${y_m}$ follows the same distribution. From Eq. (53) and Eq. (54), the following inequality
\begin{equation} \label{eq:55}
\sum\limits_{m = 1}^M {\lambda _m^2{{\left| {{x_m}} \right|}^2}}  - \sum\limits_{m = 1}^M {\lambda _m^2\left| {x_m^H{y_m}} \right|}  \gg 0
\end{equation}
holds for $M \to \infty $, then we have $\left| {{\mathbf{\hat h}}_k^H{\mathbf{\hat W}}_1^2{{{\mathbf{\hat h}}}_k}} \right| \gg \left| {{\mathbf{\hat h}}_k^H{\mathbf{\hat W}}_1^2{{{\mathbf{\hat h}}}_{k'}}} \right|$.

Similarly, we can prove that $\left| {{\mathbf{\hat h}}_k^H{\mathbf{\hat W}}_2^2{{{\mathbf{\hat h}}}_k}} \right| \!\gg\! \left| {{\mathbf{\hat h}}_k^H{\mathbf{\hat W}}_2^2{{{\mathbf{\hat h}}}_{k'}}} \right|$. So far we have proved that $\left| {{{\mathbf{h}}_k^H}{\mathbf{\hat W}}_l^2{\mathbf{\hat h}}_k} \right| \!\gg\! \left| {{{\mathbf{h}}_k^H}{\mathbf{\hat W}}_l^2{\mathbf{\hat h}}_{k'}} \right|$ for $l \in \left\{ {1,2} \right\}$, the proof for Theorem 2 is complete.

\section{Proof of Proposition 1}
For any $M, K \to \infty $ and $M \gg K$, where $k \in \left\{ {1, \cdots ,K} \right\}$ and $l \in \left\{ {1,2} \right\}$, let's derive the deterministic equivalent of ${\mathbf{h}}_k^H{{\mathbf{\hat W}}_l}{{\mathbf{\hat h}}_k}$:
\begin{eqnarray} \label{eq:56}
  &&\!\!\!\!\!\!\!\!\!\!\!\!\!\! {\mathbf{h}}_k^H{{{\mathbf{\hat W}}}_l}{{{\mathbf{\hat h}}}_k} = {\mathbf{h}}_k^H{\left( {{{{\mathbf{\hat h}}}_k}{\mathbf{\hat h}}_k^H + \sum\limits_{j = 1,j \ne k}^K {{{{\mathbf{\hat h}}}_j}{\mathbf{\hat h}}_j^H}  + M{\alpha _l}{{\mathbf{I}}_M}} \right)^{ - 1}}{{{\mathbf{\hat h}}}_k} \hfill \nonumber \\
  &&\!\!\!\!\!\!\!\!\!\!\!\!\!\! \xrightarrow{{M \to \infty }}  {\text{tr}}\left[ {{{{\mathbf{\hat h}}}_k}{\mathbf{h}}_k^H{{\left( {{{{\mathbf{\hat h}}}_k}{\mathbf{\hat h}}_k^H + M{\alpha _l}{{\mathbf{I}}_M}} \right)}^{ - 1}}} \right] \hfill \nonumber \\
  &&\!\!\!\!\!\!\!\!\!\!\!\!\!\! \xrightarrow{{M \to \infty }}  \frac{{{\mathbf{h}}_k^H{{\left( {M{\alpha _l}{{\mathbf{I}}_M}} \right)}^{ - 1}}{{{\mathbf{\hat h}}}_k}}}{{1 + {\mathbf{\hat h}}_k^H{{\left( {M{\alpha _l}{{\mathbf{I}}_M}} \right)}^{ - 1}}{{{\mathbf{\hat h}}}_k}}} \hfill \nonumber \\
  &&\!\!\!\!\!\!\!\!\!\!\!\!\!\! \xrightarrow{{M \to \infty }} \! \frac{{\frac{1}{{M{\alpha _l}}}M{\mathbf{z}}_k^H{{\left( {{\mathbf{\Theta }}_k^{{1 \mathord{\left/
 {\vphantom {1 2}} \right.
 \kern-\nulldelimiterspace} 2}}} \right)}^H}{\mathbf{\Theta }}_k^{{1 \mathord{\left/
 {\vphantom {1 2}} \right.
 \kern-\nulldelimiterspace} 2}}\left(\! {\sqrt {1 \!-\! \tau _k^2} {{\mathbf{z}}_k} + \tau {{\mathbf{q}}_k}} \!\right)}}{{\!\!1\! +\! \frac{1}{{{\alpha _l}}}\!\!\left(\! {\!\sqrt {1 \!-\! \tau _k^2} {\mathbf{z}}_k^H \!+\! \tau {\mathbf{q}}_k^H} \!\right)\!\!{{\left(\! {{\mathbf{\Theta }}_k^{{1 \mathord{\left/
 {\vphantom {1 2}} \right.
 \kern-\nulldelimiterspace} 2}}} \!\right)}^{\!\!H}}\!\!{\mathbf{\Theta }}_k^{{1 \mathord{\left/
 {\vphantom {1 2}} \right.
 \kern-\nulldelimiterspace} 2}}\!\!\left( {\!\!\sqrt {1 \!-\! \tau _k^2} {{\mathbf{z}}_k} \!+\! \tau {{\mathbf{q}}_k}} \!\!\right)}} \hfill \nonumber \\
  &&\!\!\!\!\!\!\!\!\!\!\!\!\!\!  \xrightarrow {{M \to \infty }}  \frac{{\sqrt {1 - \tau _k^2} {\text{tr}}{{\left( {{\mathbf{\Theta }}_k^{{1 \mathord{\left/
 {\vphantom {1 2}} \right.
 \kern-\nulldelimiterspace} 2}}} \right)}^H}{\mathbf{\Theta }}_k^{{1 \mathord{\left/
 {\vphantom {1 2}} \right.
 \kern-\nulldelimiterspace} 2}}}}{{M{\alpha _l} + {\text{tr}}{{\left( {{\mathbf{\Theta }}_k^{{1 \mathord{\left/
 {\vphantom {1 2}} \right.
 \kern-\nulldelimiterspace} 2}}} \right)}^H}{\mathbf{\Theta }}_k^{{1 \mathord{\left/
 {\vphantom {1 2}} \right.
 \kern-\nulldelimiterspace} 2}}}} \hfill
\end{eqnarray}
Then we have
\begin{equation} \label{eq:57}
{\mathbf{h}}_k^H{{\mathbf{\hat W}}_l}{{\mathbf{\hat h}}_k} - \frac{{\sqrt {1 - \tau _k^2} {\text{tr}}{{\left( {{\mathbf{\Theta }}_k^{{1 \mathord{\left/
 {\vphantom {1 2}} \right.
 \kern-\nulldelimiterspace} 2}}} \right)}^H}{\mathbf{\Theta }}_k^{{1 \mathord{\left/
 {\vphantom {1 2}} \right.
 \kern-\nulldelimiterspace} 2}}}}{{M{\alpha _l} + {\text{tr}}{{\left( {{\mathbf{\Theta }}_k^{{1 \mathord{\left/
 {\vphantom {1 2}} \right.
 \kern-\nulldelimiterspace} 2}}} \right)}^H}{\mathbf{\Theta }}_k^{{1 \mathord{\left/
 {\vphantom {1 2}} \right.
 \kern-\nulldelimiterspace} 2}}}}\xrightarrow{{M \to \infty }}0
\end{equation}
almost surely, and the proof for Proposition 1 is complete.

\section{Important Lemmas}
\emph{Lemma 1 (Matrix Inversion Lemma): [29, Lemma 2.2]:}
Let \textbf{U} be an $N \times N$ invertible matrix and ${\mathbf{x}} \in {\mathbb{C}^N}$, $c \in \mathbb{C}$ for which ${\mathbf{U}} + c{\mathbf{x}}{{\mathbf{x}}^H}$ is invertible. Then
\begin{equation} \label{eq:58}
{{\mathbf{x}}^H}{\left( {{\mathbf{U}} + c{\mathbf{x}}{{\mathbf{x}}^H}} \right)^{ - 1}} = \frac{{{{\mathbf{x}}^H}{{\mathbf{U}}^{ - 1}}}}{{1 + c{{\mathbf{x}}^H}{{\mathbf{U}}^{ - 1}}{\mathbf{x}}}}.
\end{equation}

\emph{Lemma 2 (Resolvent Identity):}
Let \textbf{U} and \textbf{V} be two invertible complex matrices of size $N \times N$. Then
\begin{equation} \label{eq:59}
{{\mathbf{U}}^{ - 1}} - {{\mathbf{V}}^{ - 1}} =  - {{\mathbf{U}}^{ - 1}}\left( {{\mathbf{U}} - {\mathbf{V}}} \right){{\mathbf{V}}^{ - 1}}.
\end{equation}

\emph{Lemma 3 [30, Lemma 14.2]:}
Let ${{\mathbf{A}}_1},{{\mathbf{A}}_2}, \cdots ,$ with ${{\mathbf{A}}_N} \in {\mathbb{C}^{N \times N}}$ be a series of random matrices generated by the probability space $\left( {\Omega ,\mathcal{F},P} \right)$ such that, for $w \in {\mathbf{A}} \subset \Omega $, with $P\left( A \right) = 1$, $\left\| {{{\mathbf{A}}_N}\left( w \right)} \right\| < K\left( w \right)$, uniformly on \emph{N}. Let ${{\mathbf{x}}_1},{{\mathbf{x}}_2}, \cdots ,$ with ${{\mathbf{x}}_N} \in {\mathbb{C}^N}$, be random vectors of i.i.d. entries with zero mean, variance ${1 \mathord{\left/
 {\vphantom {1 N}} \right.
 \kern-\nulldelimiterspace} N}$, and eighth-order moment of order $O\left( {{1 \mathord{\left/
 {\vphantom {1 {{N^4}}}} \right.
 \kern-\nulldelimiterspace} {{N^4}}}} \right)$, independent of ${{\mathbf{A}}_N}$. Then
\begin{equation} \label{eq:60}
{\mathbf{x}}_N^H{{\mathbf{A}}_N}{{\mathbf{x}}_N} - \frac{1}{N}{\text{tr}}{{\mathbf{A}}_N}\xrightarrow{{N \to \infty }}0
\end{equation}
almost surely.

\emph{Lemma 4:}
Let ${{\mathbf{A}}_N}$ be as in Lemma 3 and ${{\mathbf{x}}_N},{{\mathbf{y}}_N} \in {\mathbb{C}^N}$ be random, mutually independent with standard i.i.d. entries of zero mean, variance ${1 \mathord{\left/
 {\vphantom {1 N}} \right.
 \kern-\nulldelimiterspace} N}$, and eighth-order moment of order $O\left( {{1 \mathord{\left/
 {\vphantom {1 {{N^4}}}} \right.
 \kern-\nulldelimiterspace} {{N^4}}}} \right)$, independent of ${{\mathbf{A}}_N}$. Then
\begin{equation} \label{eq:61}
{\mathbf{y}}_N^H{{\mathbf{A}}_N}{{\mathbf{x}}_N}\xrightarrow{{N \to \infty }}0
\end{equation}
almost surely.

\emph{Lemma 5: [30, Lemma 14.3]:}
Let ${{\mathbf{A}}_1},{{\mathbf{A}}_2}, \cdots ,$ with ${{\mathbf{A}}_N} \in {\mathbb{C}^{N \times N}}$ be deterministic with uniformly bounded spectral norm and   ${{\mathbf{B}}_1},{{\mathbf{B}}_2}, \cdots ,$ with ${{\mathbf{B}}_N} \in {\mathbb{C}^{N \times N}}$, be random Hermitian, with eigenvalues $\lambda _1^{{{\mathbf{B}}_N}} \leqslant  \cdots  \leqslant \lambda _N^{{{\mathbf{B}}_N}}$ such that, with probability 1, there exist $\varepsilon  > 0$ for $\lambda _1^{{{\mathbf{B}}_N}} > \varepsilon $ for all large \emph{N}. The for ${\mathbf{v}} \in {\mathbb{C}^N}$
\begin{equation} \label{eq:62}
\frac{1}{N}{\text{tr}}{{\mathbf{A}}_N}{\mathbf{B}}_N^{ - 1} - \frac{1}{N}{\text{tr}}{{\mathbf{A}}_N}{\left( {{{\mathbf{B}}_N} + {\mathbf{v}}{{\mathbf{v}}^H}} \right)^{ - 1}}\xrightarrow{{N \to \infty }}0
\end{equation}
almost surely, where ${\mathbf{B}}_N^{ - 1}$ and ${\left( {{{\mathbf{B}}_N} + {\mathbf{v}}{{\mathbf{v}}^H}} \right)^{ - 1}}$ exist with probability 1.

\ifCLASSOPTIONcaptionsoff
  \newpage
\fi

\end{document}